\documentclass[manuscript,screen]{acmart}

\DeclareMathOperator{\R}{\mathbb{R}}
\DeclareRobustCommand{\bbone}{\text{\usefont{U}{bbold}{m}{n}1}}
\usepackage{algorithm} 
\usepackage{algpseudocode}
\usepackage{dsfont}
\usepackage{caption}
\usepackage{subcaption}
\usepackage{multirow}
\usepackage{graphicx,wrapfig,lipsum}
\usepackage{float}
\usepackage[absolute]{textpos}
\usepackage{bbm}
\usepackage[english]{babel}
\usepackage{pgfplots}
\usepackage[colorinlistoftodos]{todonotes}
\usepackage{enumitem}

\usepackage{tikz}
\usepackage{adjustbox}
\usetikzlibrary{pgfplots.groupplots}
\pgfplotsset{compat=1.18}
\newcommand{\algname}[1] {{\fontfamily{cmtt}\selectfont {#1}}}

\definecolor{tolblue}{HTML}{5E81B5}
\definecolor{tolorange}{HTML}{EB6235}
\definecolor{tolgreen}{HTML}{8FB032}
\definecolor{tolpurple}{HTML}{8778B3}
\definecolor{tolyellow}{HTML}{E19C24}

\pgfplotsset{
    legend image with text/.style={
        legend image code/.code={%
            \node[anchor=center] at (0.3cm,0cm) {#1};
        }
    },
    every tick label/.append style={font=\large},
    compat=1.3,
}
\definecolor{_blue}{RGB}{86,118,167}
\definecolor{_orange}{RGB}{213,146,55}
\definecolor{_green}{RGB}{137,164,62}
\definecolor{_red}{RGB}{220,96,57}
\definecolor{_purple}{RGB}{120,111,165}
\definecolor{_cyan}{RGB}{133,148,229}
\definecolor{_yellow}{RGB}{201,171,55}
\definecolor{_gray}{RGB}{180,180,180}
\definecolor{_lightgray}{RGB}{240,240,240}

\AtBeginDocument{%
  }
\setcopyright{cc}
\copyrightyear{2026}
\acmYear{2026}
\acmDOI{}
\acmISBN{}



\begin{document}

\title{The Unfairness of Multifactorial Bias in Recommendation}

\author{Masoud Mansoury}
\email{m.mansoury@tudelft.nl}
\affiliation{%
  \institution{Delft University of Technology}
  \city{Delft}
  \country{The Netherlands}
}
\author{Jin Huang}
\email{jh2642@cam.ac.uk}
\affiliation{%
  \institution{University of Cambridge}
  \city{Cambridge}
  \country{United Kingdom}
}
\author{Mykola Pechenizkiy}
\email{m.pechenizkiy@tue.nl}
\affiliation{%
  \institution{Eindhoven University of Technology}
  \city{Eindhoven}
  \country{The Netherlands}
}
\author{Herke van Hoof}
\email{h.c.vanhoof@uva.nl}
\affiliation{%
  \institution{University of Amsterdam}
  \city{Amsterdam}
  \country{The Netherlands}
}
\author{Maarten de Rijke}
\email{m.derijke@uva.nl}
\affiliation{%
  \institution{University of Amsterdam}
  \city{Amsterdam}
  \country{The Netherlands}
}

\renewcommand{\shortauthors}{Mansoury et al.}

\begin{abstract}
Popularity bias and positivity bias are two prominent sources of bias in recommender systems. Both arise from input data, propagate through recommendation models, and lead to unfair or suboptimal outcomes. Popularity bias occurs when a small subset of items receives most interactions, while positivity bias stems from the over-representation of high rating values. Although each bias has been studied independently, their \textit{combined effect}, to which we refer to as \textit{multifactorial bias}, remains underexplored. In this work,
we examine how multifactorial bias influences item-side fairness, focusing on exposure bias, which reflects the unequal visibility of items in recommendation outputs. Through simulation studies, we find that positivity bias is disproportionately concentrated on popular items, further amplifying their over-exposure. Motivated by this insight, we adapt a percentile-based rating transformation as a pre-processing strategy to mitigate multifactorial bias. Experiments using six recommendation algorithms across four public datasets show that this approach improves exposure fairness with negligible accuracy loss. We also demonstrate that integrating this pre-processing step into post-processing fairness pipelines enhances their effectiveness and efficiency, enabling comparable or better fairness with reduced computational cost. These findings highlight the importance of addressing multifactorial bias and demonstrate the practical value of simple, data-driven pre-processing methods for improving fairness in recommender systems.
\end{abstract}

\maketitle

\section{Introduction}

\begin{figure*}[t!]
    \centering
    \begin{subfigure}[b]{0.6\textwidth}
        \includegraphics[width=\textwidth]{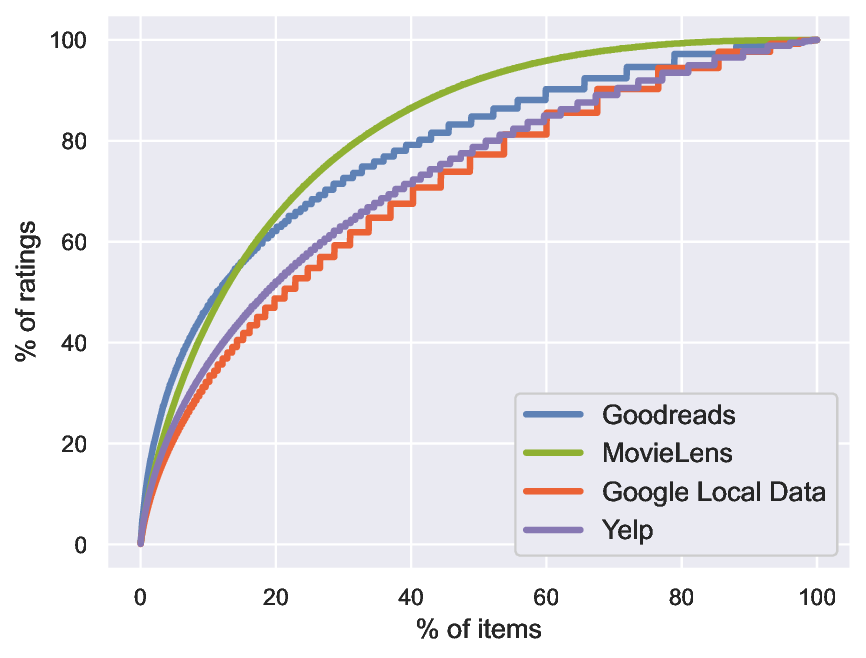}
    \end{subfigure}
\caption{Popularity distribution of the rating data in multiple datasets used in this paper. The horizontal axis shows the percentage of the items and the vertical axis shows the percentage of the accumulated ratings.}\label{fig_ml_longtail_dist}
\end{figure*}

Recommender systems are known to suffer from different types of bias against users and items~\cite{chen2023bias,steck2011item,schnabel2016recommendations,pradel2012ranking,ovaisi2020correcting,marlin2009collaborative}. Two prominent forms of such bias are \textit{popularity bias}~\cite{canamares2018should,steck2011item,pradel2012ranking,klimashevskaia2024survey,chen2023bias} and \textit{positivity bias}~\cite{pradel2012ranking,chen2023bias}.
Popularity bias refers to the situation that the majority of rating data is concentrated on a small subset of highly popular items, leading to their over-representation while less popular items remain under-represented in the input data. Figure~\ref{fig_ml_longtail_dist} shows this phenomenon across four datasets. As shown, most ratings are concentrated on a small fraction of items, with MovieLens exhibiting the strongest popularity bias. For instance, in the MovieLens dataset, approximately 20\% of the items make up around 65\% of the total ratings. The abundance of rating data on popular items skews the learning process of the recommender models, causing them to primarily learn the characteristics of these popular items~\cite{schnabel2016recommendations,huang2024going}. Consequently, these items are disproportionately recommended, reinforcing their dominance in recommendation results.

\begin{figure*}[t!]
    \centering
    \begin{subfigure}[b]{0.6\textwidth}
        \includegraphics[width=\textwidth]{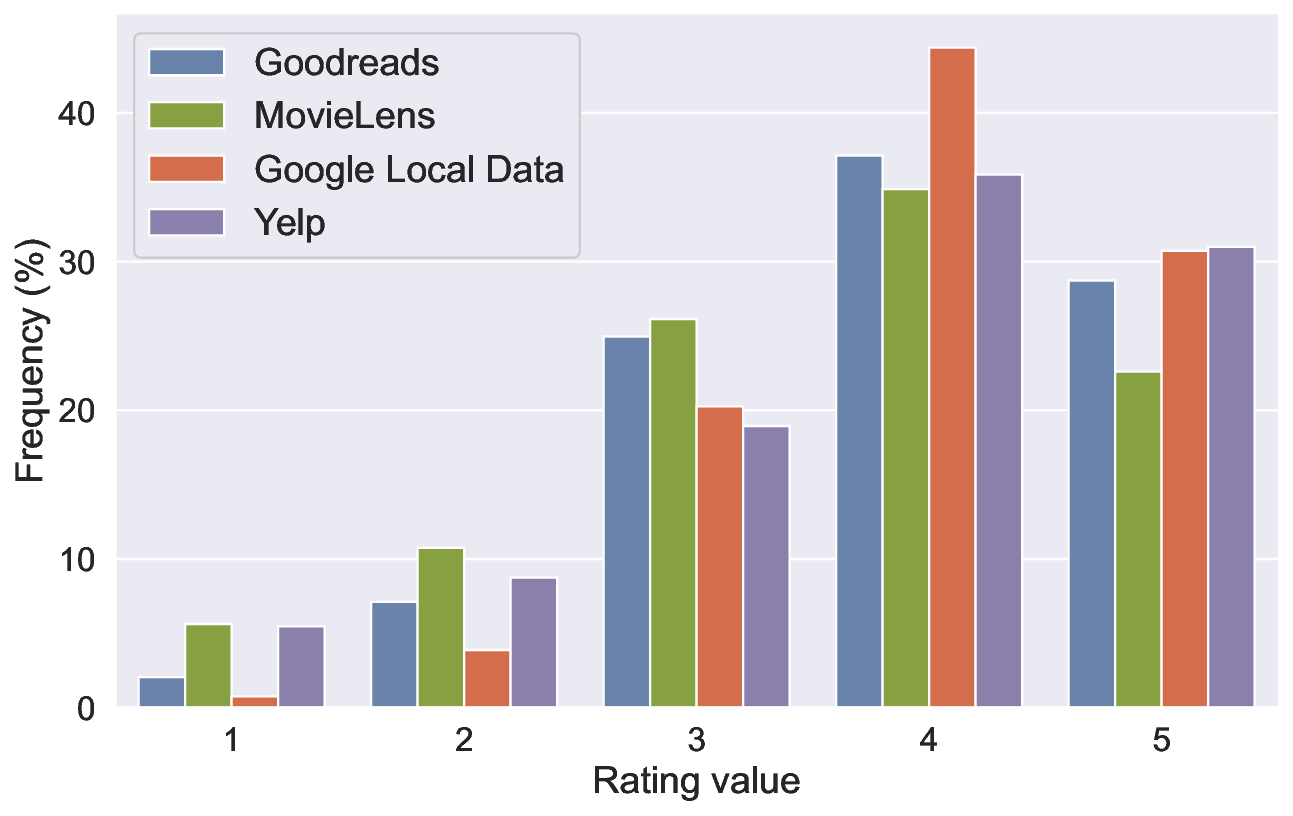}
    \end{subfigure}
\caption{Distribution of ratings of items in the MovieLens dataset. The horizontal axis shows the rating values in the dataset and the vertical axis shows the percentage of times each rating value is seen in the dataset.}
\label{fig_ml_rating_dist}
\end{figure*}

On the other hand, positivity bias refers to the situation that positive feedback (high rating values) is over-represented in the input data. This occurs because users are more likely to rate items they enjoy while ignoring those they dislike. Figure~\ref{fig_ml_rating_dist} shows the distribution of ratings across four datasets, highlighting that higher ratings are significantly more prevalent. This pattern is even stronger in the Google Local Data. 

These types of bias originate from the input data, are exacerbated or transferred by recommender algorithms, and lead to \textit{unfairness} in recommendation results~\cite{klimashevskaia2024survey,wang2023survey,jin2023survey}. Unfair recommendations not only negatively affect the recommendation performance~\cite{abdollahpouri2019unfairness,mansoury2020feedback} but also lead to or perpetuate undesirable social dynamics~\cite{singh2018fairness,li2021user,yoo2024ensuring}. One specific form of this type of unfairness in recommendation output is \textit{exposure bias}~\cite{singh2018fairness,zehlike2020reducing,mansoury2022understanding}, which refers to the situation that certain items frequently appear in recommendation results while certain other items are never or rarely recommended to users. 

Prior research has discovered various consequences of this disparate representation of items in recommendation results: (i)~it negatively affects user experience by limiting exposure to a small subset of items~\cite{mansoury2020feedback,sinha2016deconvolving,yang2023rectifying,mansoury2024mitigating} or failing to align with users' preferences~\cite{huang2024going,huang2020keeping,schnabel2016recommendations}, (ii)~it leads to supplier unfairness, as items from certain suppliers would not be sufficiently (or equally) visible in the recommendation results~\cite{mansoury2022understanding,mehrotra2018towards}, and (iii)~it raises ethical concerns in sensitive domains such as job recommendation or online dating, where the recommended items represent individuals belonging to different sensitive attribute classes, increasing the risk of discrimination~\cite{singh2018fairness,edizel2020fairecsys,mansoury2020investigating}. In this paper, we explore two key types of bias in input data, popularity bias and positivity bias, their connection to exposure bias in the recommendation results, and strategies to mitigate these types of bias to improve exposure fairness.

Although the unfairness of popularity bias and positivity bias have been \textit{separately} studied in the literature~\cite{abdollahpouri2019unfairness,kowald2020unfairness,naghiaei2022unfairness,rahmani2022unfairness,abdollahpouri2020connection,allard2020negative}, their combined effect on recommendation fairness remains an open research question. This combined effect is referred to as \textit{multifactorial bias}, which encapsulates the essential properties of both types of bias.  
\citet{huang2024going} study the effect of this bias on the performance and accuracy of recommendation models, but its impact on the fairness of recommendation models has yet to be explored. To fill this gap, in this paper, we conduct a comprehensive study on the unfairness caused by multifactorial bias in recommendation results and propose a solution to mitigate it. 

First, we investigate multifactorial bias in several publicly-available datasets. Our analysis reveals that positivity bias is stronger among popular items, signifying that highly rated feedback is disproportionately concentrated on popular items, further reinforcing their over-representation in recommendations. We then conduct a simulation study to examine how this type of bias affects recommendation performance and fairness. Our results show that mitigating positivity bias, particularly among popular items (which remain popular in terms of number of interactions/ratings), can improve fairness of exposure for items in recommendation results.

Next, we adapt an existing rating transformation method~\cite{mansoury2021flatter} to address multifactorial bias in rating data, specifically by diminishing the impact of positivity bias on popular items. This method transforms the raw rating data into percentile values in item profiles, leading to less skew in percentile distribution in the input data. Through extensive experiments using six recommendation algorithms across four datasets, we validate the effectiveness of this proposed method in enhancing exposure fairness of recommendation results. 

It is important to note that the primary objective of this work is not to introduce a new bias mitigation algorithm. Rather, our aim is to demonstrate how an existing method that balances rating distributions can be effectively adapted to address multifactorial bias, particularly the concentration of positivity bias among popular items, which is a property shared by both the proposed method and the bias itself.

Finally, beyond improving fairness, we show that our proposed method enhances the efficiency of existing post-processing bias mitigation approaches. The effectiveness of these approaches depends heavily on the size of the initial recommendation lists generated by the recommendation model. Longer lists yield fairer recommendations, but at the expense of higher computational costs. We show that our pre-processing approach enables achieving fairer recommendation results with much smaller initial recommendation lists, resulting in more efficient fair recommendation pipeline. 

Overall, we make the following contributions:

\begin{itemize}
    \item We investigate multifactorial bias in publicly-available recommendation datasets, revealing a strong correlation between positivity bias and popularity bias: positivity bias is more prevalent on popular items.
    \item Through a simulation study, we demonstrate the relationship between multifactorial bias and exposure fairness in recommendation results, showing that this type of bias negatively affects item exposure fairness.
    \item We adapt an existing rating transformation method to mitigate the effect of multifactorial bias in recommendation datasets. Our analysis shows that this proposed method reduces the effect of positivity bias, particularly on popular items, resulting in reduced multifactorial bias in input datasets and fairer item exposure in recommendation results.
    \item Lastly, through additional sets of experiments, we show that the proposed pre-processing method improves the efficiency of existing post-processing bias mitigation methods by achieving comparable performance with substantially shorter initial recommendation list as input for those post-processing bias mitigation methods.
\end{itemize}
\section{Related Work}\label{sec_relatedwork}

\subsection{Bias and unfairness}

\subsubsection{Types of bias}
Selection bias, which refers to bias present in the data associated with recommender systems, is pervasive in recommendation applications and manifests in various forms~\citep{chen2023bias}.
It can arise due to self-selection bias, with users choosing to interact with certain items more often~\citep{pradel2012ranking,steck2011item}. Or it may be due to algorithmic bias: the recommender system used for logging choosing to show certain items to users more often~\citep{baeza2018bias,schnabel2016recommendations}.
Among different types of bias, popularity bias and positivity bias are the most widely recognized and studied, manifesting as the over-representation of interactions with popular items~\cite{canamares2018should,pradel2012ranking,steck2011item} or items that users prefer~\citep{pradel2012ranking}, respectively.
Additionally, two forms of self-selection bias are frequently observed: incentive bias, where users are incentivized to
provide ratings for benefits and rewards~\citep{panniello2016impact}, and conformity bias, where users tend to rate items similarly to others in a group~\citep{knyazev2022bandwagon,krishnan2014methodology}. 
These types of bias are often characterized by a single underlying factor. 

In practice, selection bias is typically more complex, often involving a combination of multiple biases or reflecting a multifactorial structure determined by more than one factor~\citep{zheng2022cbr,wu2021unbiased}.
Selection bias can also be influenced by the additional factor of time, resulting in dynamic popularity bias~\citep{huang2022different}.
Moreover, contextual factors such as position, modality, or surrounding items in a recommendation list can simultaneously affect user interactions, referred to as surrounding item bias or contextual bias~\citep{wu2021unbiased,zhuang2021cross,sarvi-2023-impact}.
Additionally, selection correlated with both item and rating value factors has been observed across multiple real-world datasets~\citep{pradel2012ranking,huang2020keeping,huang2024going}.
Specifically, \citet{huang2024going} formally define a multifactorial bias that is determined by two factors, viz.\ item and rating value, which can be seen as a generalization of popularity and positivity bias.
In this paper, we build on the findings from~\citet{huang2024going} and focus on multifactorial bias and its resulting unfairness problem.

\subsubsection{Unfairness}

Despite being two distinct concepts, bias and unfairness are closely interrelated. 
A widely accepted view is that various forms of bias and the learning process are the key reasons for unfairness issues in machine learning~\citep{li2023fairness,abdollahpouri2019unfairness,wang2023survey}.
For instance, prior studies have shown that popularity bias can lead to unfair treatment of both long-tail items and users with little interest in popular items~\citep{yalcin2022evaluating,abdollahpouri2019unfairness}.
Recently, \citet{dai2024bias} offer a unified perspective by framing both bias and unfairness as instances of distribution mismatch -- bias arises from a mismatch with the objective target distribution in real scenarios, while unfairness reflects a mismatch with the subjective distribution aligned with human values.

The definition of fairness in recommender systems can vary depending on the multi-stakeholder settings, commonly encompassing user-oriented~\citep{li2021user,abdollahpouri2019unfairness,li2023fairness} or item-oriented fairness perspectives~\citep{li2023fairness,abdollahpouri2019multi,singh2019policy}.
User fairness typically focuses on avoiding algorithmic discrimination against certain users that are defined by demographic attributes (e.g., gender and age)~\citep{zhu2018fairness,steck2018calibrated}.
In contrast, item fairness aims to ensure that items from different categories (e.g., long-tail items) receive equitable exposure in the recommendation results~\citep{mansoury2020fairmatch}.
Each of these fairness perspectives can be considered at both the individual~\citep{steck2018calibrated,yao2017beyond} and group levels~\citep{zehlike2020reducing,zehlike2017fa,singh2018fairness}.
At the individual level, fairness is often defined as treating similar individuals similarly, whereas at the group level, fairness refers to ensuring that different user or item groups receive equitable treatment or outcomes, such as equal quality of recommendations or balanced exposure~\citep{mehrabi2021survey,castelnovo2022clarification}.
For example, \citet{steck2018calibrated} introduces calibration fairness, an individual-level user fairness notion, which requires that users receive recommendations that are close to what they are expecting to get based on the type of items they have rated.
Rather than focusing on one specific fairness consideration, the concept of hybrid fairness acknowledges the diverse fairness demands that arise in the real world and aims to achieve more than one fairness requirement simultaneously~\citep{abdollahpouri2019multi,mehrotra2018towards,mansoury2022understanding}.

\subsection{Fair recommendation methods}
Existing methods for improving fairness in recommender systems can be broadly categorized based on their position in the recommendation pipeline: data-oriented, ranking, and re-ranking methods, which are also referred to as pre-processing, in-processing, and post-processing methods, respectively.

Data-oriented methods mitigate unfairness by modifying or augmenting the training data used to train the recommendation model.
For example, \citet{ekstrand2018all} apply a re-sampling strategy to adjust the proportion of different user groups in the training data, thereby addressing user fairness issues arising from imbalanced user group distributions.
\citet{rastegarpanah2019fighting} propose the use of antidote data (e.g., fake user data) added to the training data to counteract unfairness.

Ranking methods incorporate fairness considerations directly into the learning process of recommendation methods.
These methods are typically categorized into regularization-based methods~\citep{yao2017beyond,zehlike2020reducing,burke2018balanced,beutel2019fairness}, adversarial learning-based methods~\citep{bose2019compositional,wu2021learning}, and reinforcement
learning-based methods~\citep{ge2021towards}.

In contrast, re-ranking methods adjust the outputs of RSs to promote fairness without modifying or retraining the underlying recommendation methods.
\citet{zehlike2017fa} propose FA*IR, a top-$k$ ranking method that maximizes utility while satisfying ranked group fairness.
\citet{sonboli2020opportunistic} define a personalized fairness score based on multiple item attributes and achieve a tradeoff between fairness and utility.
Several other works propose diversification re-ranking approaches to increase the representation of less popular items while maintaining acceptable recommendation accuracy~\citep{abdollahpouri2019managing,antikacioglu2017post,mansoury2020fairmatch}. 
Additionally, \citet{liu2024interact} examine model-generated explanations using a causal graph to identify and mitigate sources of bias in the recommendation process. They simulate user feedback on these explanations to better understand and correct biased selection mechanisms. Similarly, inspired by causal inference principles, \citet{zhang2023recommendation} analyze both item-level and feature-level popularity bias in interaction data and propose separate inverse propensity score (IPS)-based methods to mitigate their effects.

\section{Notation and Preliminaries}

We consider a typical recommendation scenario, such as movie, music, and product recommendations, where users are allowed to express their opinions on various items through explicit ratings. These ratings are then used to train a recommendation model that generates a recommendation list for each user. 

Formally, we denote $\mathcal{U}=\{u_1,\ldots,u_n\}$ as the set of $n$ users, $\mathcal{I}=\{i_1,\ldots,i_m\}$ as the set of $m$ items, and $R \in \R^{n \times m}$ as the user-item rating matrix where entry $R_{ui}$ represents the rating provided by user $u \in \mathcal{U}$ on item $i \in \mathcal{I}$. The rating values are chosen from a predefined set of numerical values within the range $[a,b]$, such that $R_{ui}\in\{r_1,r_2,\ldots,r_O\}$, where $O$ is the number of distinct rating levels. For example, in the MovieLens dataset, a 5-star rating system is used with $a=1$, $b=5$, and $R_{ui}\in\{1,2,3,4,5\}$. 

We divide items into two groups based on their popularity: head and tail. Head items denoted by $\mathcal{I}^H$ consist of popular items that make up roughly 20\% of the ratings according to the Pareto Principle \cite{sanders1987pareto}. Tail items denoted by $\mathcal{I}^T$ consist of the rest of the less popular items (the ones not in $\mathcal{I}^H$).

Given the rating matrix $R$ as input, the recommendation problem we consider is defined as building a recommendation model that effectively identifies $K$ most relevant items for a target user. The recommendation list for a user $u$ is denoted as $L_u$, which is a ranked list of $K$ items sorted in descending order based on their predicted relevance score. The set of recommendation lists for all users is denoted as $L$. 

\begin{figure*}[t!]
    \centering
    \begin{subfigure}[b]{0.42\textwidth}
        \includegraphics[width=\textwidth]{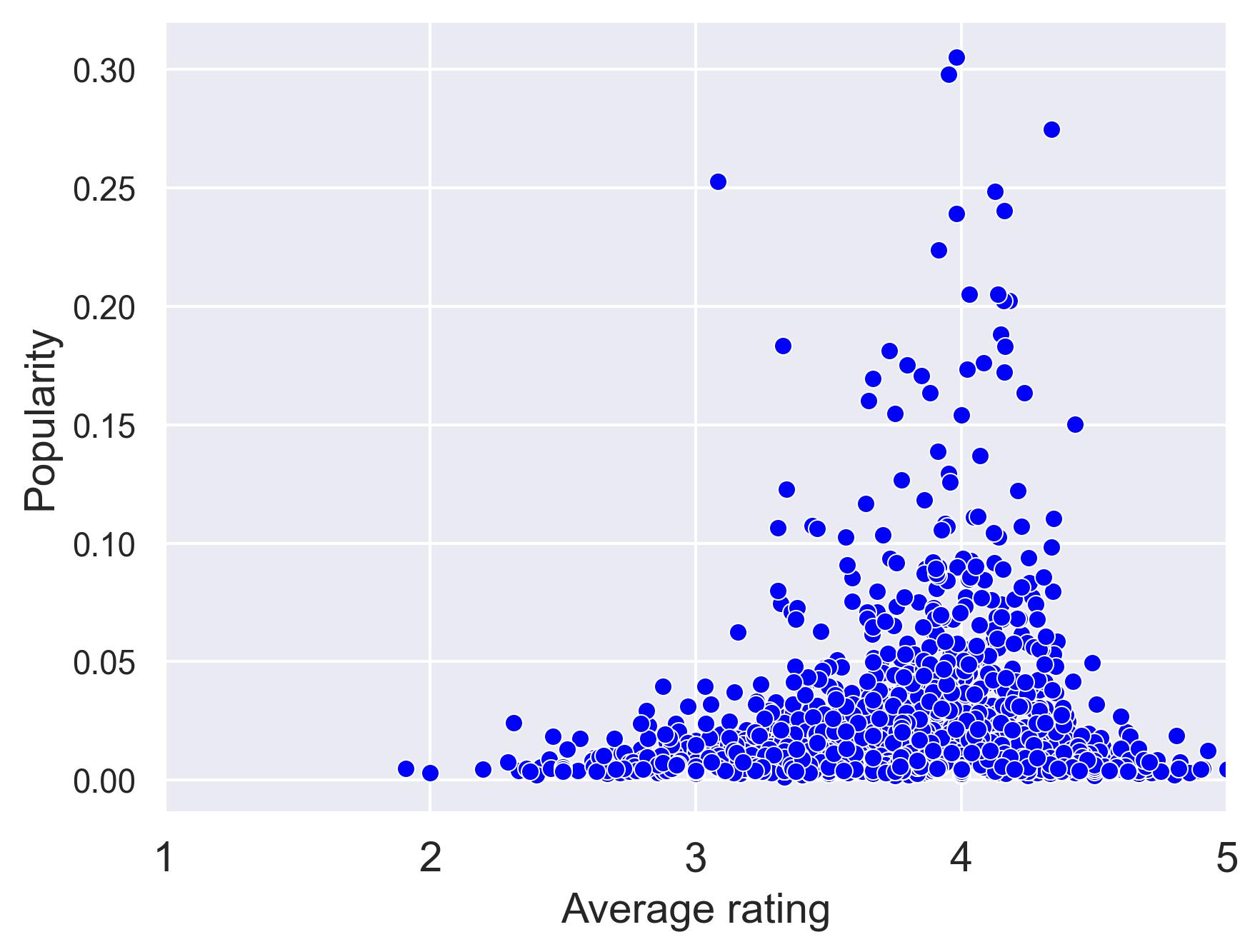}
        \caption{Goodreads}\label{fig_goodreads_pop_rating}
    \end{subfigure}
    \begin{subfigure}[b]{0.42\textwidth}
        \includegraphics[width=\textwidth]{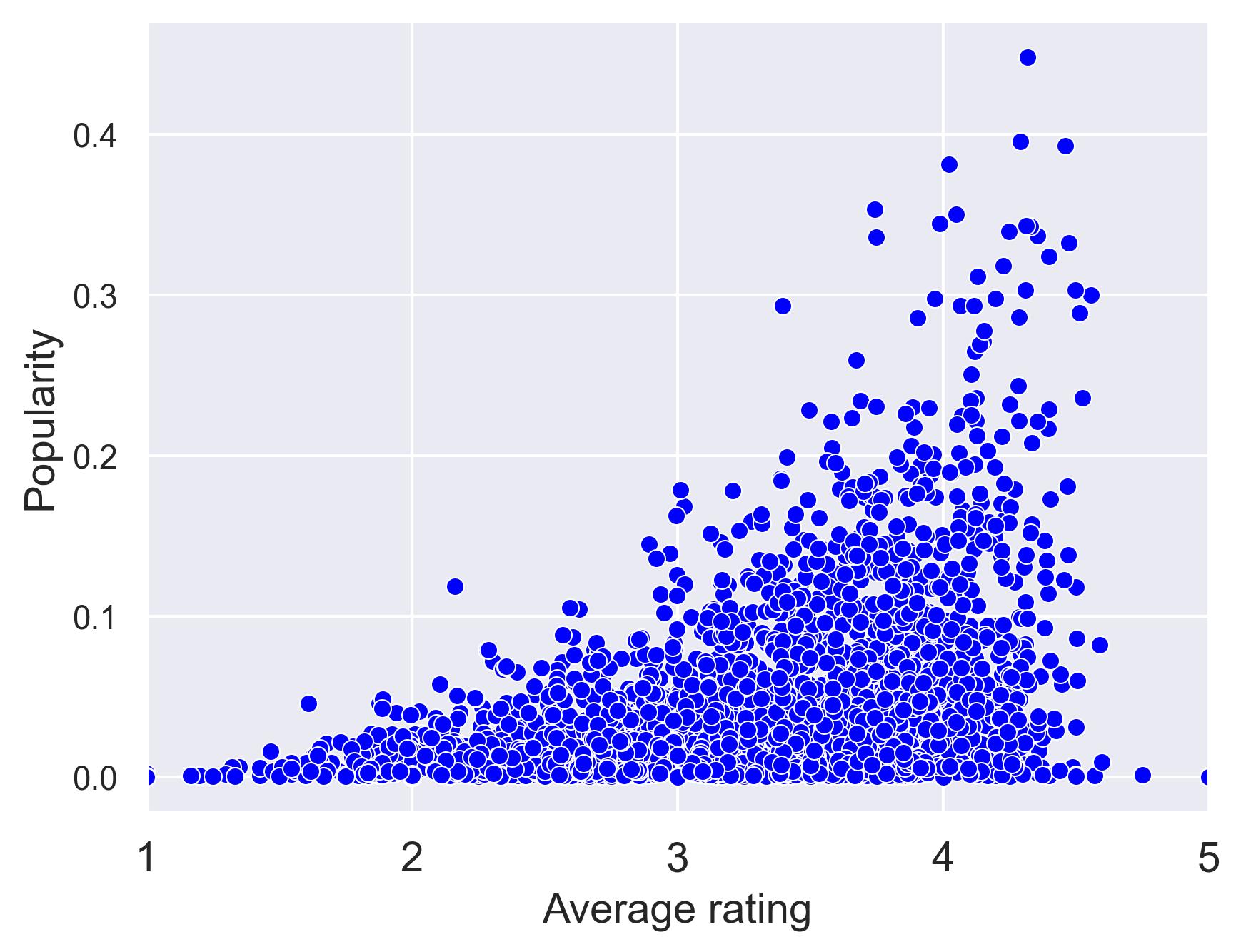}
        \caption{MovieLens}\label{fig_ml_pop_rating}
    \end{subfigure}
    \begin{subfigure}[b]{0.42\textwidth}
        \includegraphics[width=\textwidth]{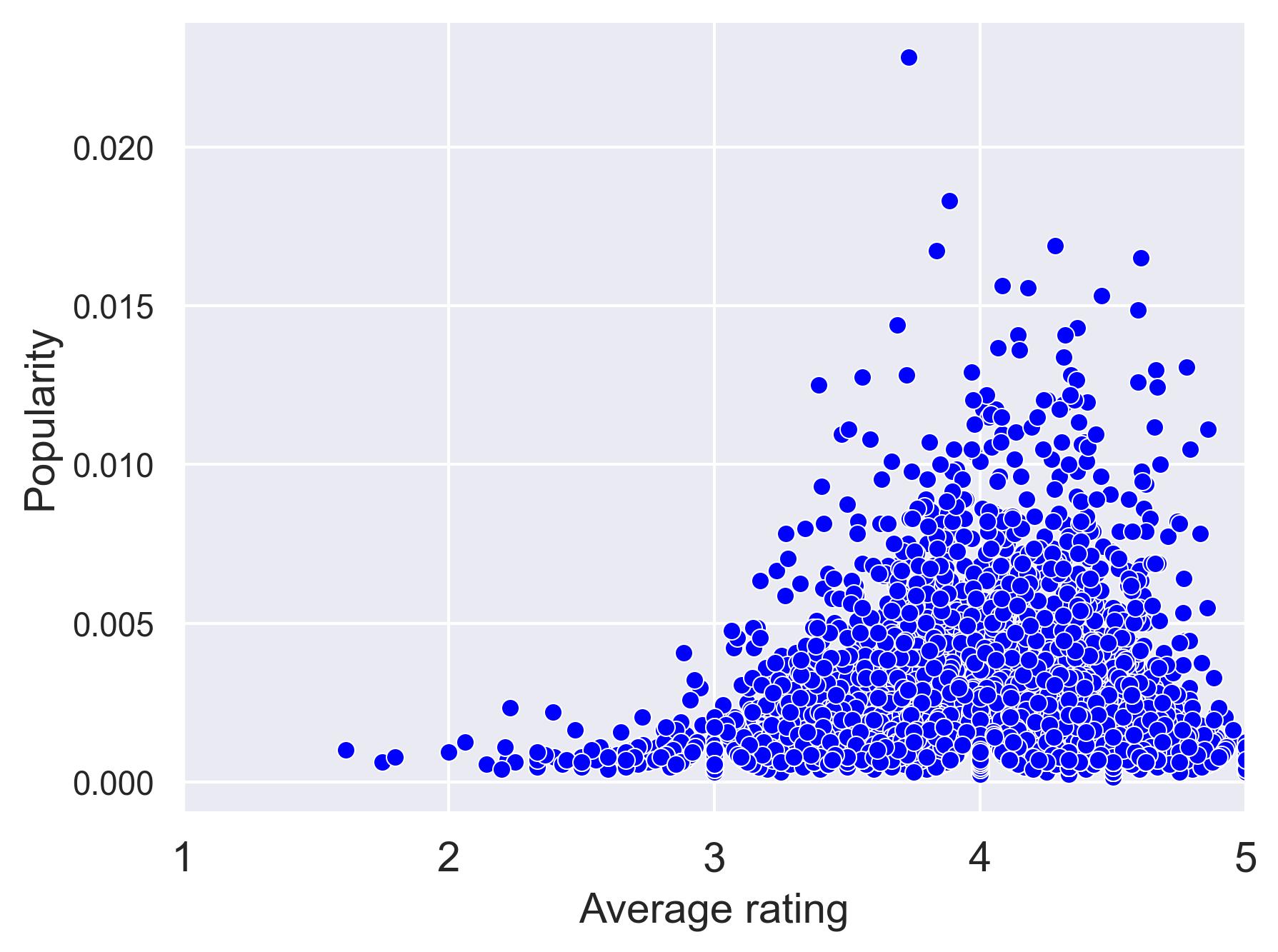}
        \caption{Google Local Data}\label{fig_glr_pop_rating}
    \end{subfigure}
    \begin{subfigure}[b]{0.42\textwidth}
        \includegraphics[width=\textwidth]{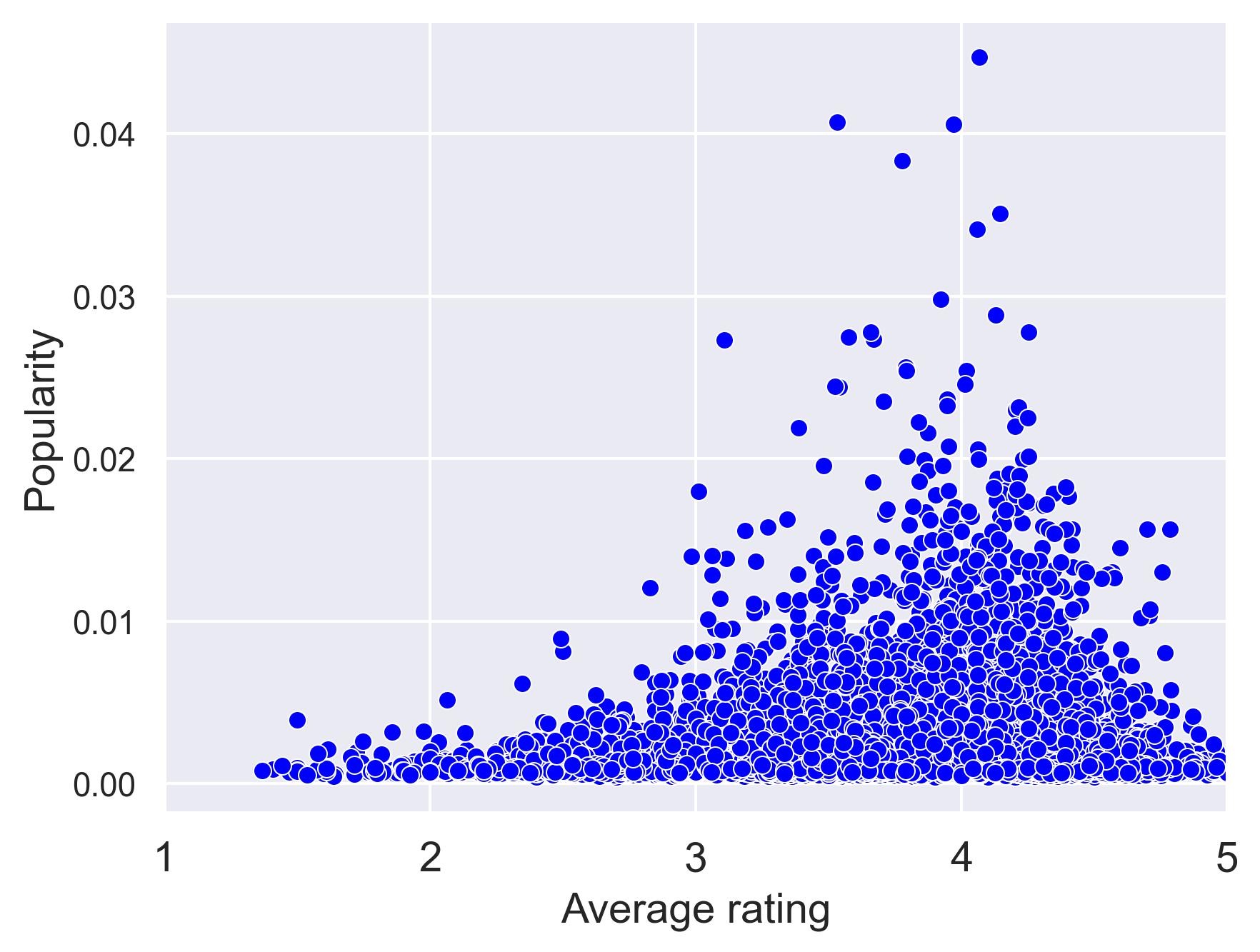}
        \caption{Yelp}\label{fig_yelp_pop_rating}
    \end{subfigure}
\caption{The relationship between average ratings and popularity of items in four public datasets.}\label{fig_pop_rating}
\end{figure*}

\subsection{Multifactorial bias}

According to~\citep{chen2023bias,huang2024going,steck2011item}, a typical recommender system with explicit user ratings often exhibits three key characteristics:
\begin{enumerate}[label=(\textbf{Characteristic \arabic*}),leftmargin=*,ref=Characteristic~\arabic*]
    \item Few popular items are frequently interacted with, while the majority of unpopular items are rarely interacted with. \label{char:pop}
    \item Users tend to rate items they like, resulting in a skewed distribution of observed ratings toward higher values. \label{char:pos}
    \item The majority of high rating values are assigned to the popular items. \label{char:mul}
\end{enumerate}
Figure~\ref{fig_pop_rating} illustrates the relationship between item popularity and average rating values across four public datasets, confirming the presence of \ref{char:mul}.

These characteristics are commonly recognized as forms of selection bias in recommender systems.
\ref{char:pop} and \ref{char:pos} are often referred to as \emph{popularity bias} and \emph{positivity bias}, respectively, which have been studied extensively in the literature, with various methods proposed to mitigate them~\citep{chen2023bias,steck2011item,pradel2012ranking,canamares2018should}. 
\ref{char:mul} has also been observed by~\citet{huang2024going}.
Accordingly, \citet{huang2024going} define a notion of \emph{multifactorial bias} that occurs if the process that decides whether a user
provides a rating is not a random selection and is determined by
the item and rating value factors. 
This definition provides a general conceptual framework, whereas \ref{char:mul} reflects a specific real-world phenomenon commonly observed in practice. 

While \ref{char:mul} more accurately reflects real-world user behavior, it remains underexplored in the literature.
\citet{huang2024going} are the first to examine the effect of multifactorial bias on the performance and accuracy of recommendation models.
In this paper, we further investigate its impact on the fairness of recommendation models.

\section{Methodology}

In this section, we describe the datasets, recommendation algorithms, evaluation metrics, and the experimental setup that we use in the rest of the paper to perform our experiments.

\captionsetup[table]{skip=4pt}
\begin{table}[t]
\aboverulesep=0ex 
\belowrulesep=0ex 
\centering
\setlength{\tabcolsep}{3pt}
\captionof{table}{Characteristics of the datasets after pre-processing.} \label{tab_dataset}
\begin{tabular}{l rrrrc}
\toprule
 Dataset & \#users & \#items & \#ratings & Density & Ratings \\
 \midrule

 Goodreads & 2,225 & 3,423 & 137,045 & 1.8\% & 1--5 \\
 MovieLens & 6,040 & 3,706 & 1,000,209 & 4.47\% & 1--5 \\
 Google Local Data & 12,791 & 16,101 & 391,715 & 0.19\% & 1--5 \\
 Yelp & 24,372 & 18,569 & 1,160,487 & 0.26\% & 1--5 \\

\bottomrule
\end{tabular}
\end{table}

\subsection{Datasets}\label{sec_dataset}

We use the following datasets for our experiments in this paper.

\begin{itemize}
    \item \textbf{Goodreads}\footnote{\url{https://github.com/BahramJannesar/GoodreadsBookDataset}}: A book recommendation dataset that contains user ratings of books. 
    \item \textbf{MovieLens}\footnote{\url{https://grouplens.org/datasets/movielens/1m/}} \cite{harper2015movielens}: A movie recommendation dataset that contains user ratings of movies.
    \item \textbf{Google Local Data}\footnote{\url{https://jiachengli1995.github.io/google/index.html}} \cite{li2022uctopic}: A subset of the Google Local Data, which consists of business reviews collected from Google Maps that contain user ratings of businesses.
    \item \textbf{Yelp}\footnote{\url{https://www.yelp.com/dataset}}: A restaurant recommendation dataset that contains user ratings of restaurants.
\end{itemize}

\noindent%
To alleviate the sparsity issue in the datasets and for more robust evaluation, we perform the following pre-processing on each dataset. From the Goodreads and Google Local Data datasets, we extract users who provided at least 10 ratings and items that received at least 10 ratings. From the Yelp dataset, we create a sample of the dataset in which each user has rated at least 20 items and each item has received at least 20 ratings. Table~\ref{tab_dataset} summarizes the characteristics of the pre-processed datasets that we use in the paper.

\subsection{Recommendation algorithms}\label{sec_algs}

Since we are studying the impact of positivity bias in recommender systems in this paper, to be able to properly analyze the impact of our proposed approaches, we only focus on recommendation algorithms that use rating data for their operation. Algorithms operating on implicit or binary ratings data (e.g., Bayesian Personalized Ranking \cite{rendle2009bpr}) would not be affected by positivity bias as they do not use rating data for their internal processing. Therefore, we perform our experiments using the following recommendation algorithms.

\begin{itemize}
    \item \textbf{Biased Matrix Factorization} (\algname{BiasedMF}) \cite{koren2009matrix}: This algorithm extends matrix factorization by incorporating the user bias and the item bias terms into the objective function to better capture the differences in users' behavior and items' characteristics.
    \item \textbf{Integrated Latent Factor and Neighborhood Models} (\algname{SVD++}) \cite{koren2008factorization}: This algorithm extends matrix factorization by incorporating neighborhood influence into the objective function.
    \item \textbf{Weighted Matrix Factorization} (\algname{WRMF}) \cite{hu2008collaborative}: A weighted matrix factorization algorithm that incorporates a weight into the objective function as the confidence in users' interest toward items. 
    \item \textbf{Listwise Matrix Factorization} (\algname{ListRank})~\cite{shi2010list}: A specific form of matrix factorization with list-wise optimization process.
    \item \textbf{User-based Neighborhood Model} (\algname{UserKNN}) \cite{resnick1994grouplens}: A nearest neighbor-based model that uses the user-user similarity matrix to recommend items based on the opinions of similar users.
    \item \textbf{Item-based Neighborhood Model} (\algname{ItemKNN}) \cite{sarwar2001item}: A nearest neighbor-based model that uses the item-item similarity matrix to recommend items similar to the ones in a target user's profile.
\end{itemize}

\subsection{Metrics}\label{sec_metrics}

We evaluate the performance of the recommendation models on accuracy and fairness metrics. We use the following metrics to evaluate the accuracy of the recommendation models.

\begin{itemize}
    \item \textbf{Precision}: The fraction of correctly recommended items (the items in the recommendation lists of users that also exist in the test data). 
    \item \textbf{Normalized Discounted Cumulative Gain (nDCG)}: A measure of ranking quality of the recommendations. nDCG weights the correctly recommended items based on their position in the recommendation list, recommending these items in higher position would lead to higher weight (higher nDCG). 
\end{itemize}

Moreover, we use the following metrics to evaluate the fairness of the recommendation results.

\begin{itemize}
    \item \textbf{Item Aggregate Diversity (IA)~\citep{mansoury2021graph}}: The fraction of items that appear at least $\alpha$ times in the recommendation lists and can be calculated as:
    \begin{equation}
        IA(L)=\frac{\sum_{i \in \mathcal{I}}{\bbone\left(\left[\sum_{u \in \mathcal{U}}{\sum_{j \in L_u}}{\bbone(i=j)}\right]\geq \alpha\right)}}{|\mathcal{I}|}.
    \end{equation}
    \noindent This metric is a generalization of standard aggregate diversity as it is used in other work \cite{adomavicius2011maximizing,vargas2011rank} where $\alpha=1$.
    
    \item \textbf{Long-tail Item Aggregate Diversity (LIA)~\citep{mansoury2021graph}}: The fraction of long-tail items that appear ar least $\alpha$ times in the recommendation lists and can be calculated as:
    \begin{equation}
        LIA(L)=\frac{\sum_{i \in \mathcal{I}^T}{\bbone\left(\left[\sum_{u \in \mathcal{U}}{\sum_{j \in L_u}}{\bbone(i=j)}\right]\geq \alpha\right)}}{|\mathcal{I}^T|}.
    \end{equation}
    
    \item \textbf{Equality of Exposure (EE}~\citep{antikacioglu2017post}): A measure of fair distribution of recommended items. It takes into account how uniformly items appear in recommendation lists. Given $E(i|L)$ as the exposure of item $i$ in the recommendation lists $L$ calculated as:
    \begin{equation}
        E(i|L) = \frac{\sum_{u \in \mathcal{U}}{\sum_{j \in L_u}{\bbone(i=j)}}}{|\mathcal{U}| \times |K|},
    \end{equation}
    \noindent we first form the distribution of items exposure as $\{E(i|L)\mid \forall i \in \mathcal{I}\}$. Then, we calculate the Gini Index over this distribution to measure its uniformity as follows:
    \begin{equation}
        \mathit{Gini}(L)=\frac{1}{|\mathcal{I}|-1}\sum_{k=1}^{|\mathcal{I}|}{(2k-|\mathcal{I}|-1)E(i_k|L)},
    \end{equation}
    \noindent where $E(i_k|L)$ is the exposure of the $k$-th least recommended item being drawn from $L$. A uniform distribution will have a Gini Index equal to zero, which is the ideal case (a lower Gini index is better). Therefore, to be consistent with the metrics definition, we define equality of exposure as $EE=1-\mathit{Gini}(L)$. This way, an $EE$ value closer to 1 signifies fairer recommendation results and is more desired (and $EE$ closer to 0 is the least fair recommendation results).
\end{itemize}

\subsection{Experimental setup}

For our experiments, we split the input data into 80\% as training set and 20\% as test set. The split is performed on users' individual profiles: for each user, we randomly retrieve and use 80\% of the ratings in her profile to form the training set, and we use the rest of the ratings to form the test set. The training set is used for building the recommendation model and generating the recommendation list for each user. The test set is used to evaluate the generated recommendations.

In the training phase, we perform extensive experiments with different hyperparameter values for each algorithm and dataset combinations to find the best-performing results. Hence, we perform a grid search over the hyperparameter space for each algorithm. For the neighborhood models (\algname{UserKNN} and \algname{ItemKNN}), we set the \textit{shrinkage rate} as $\{50,100,200\}$ and the number of neighbors as $\{10,20,30,50,70,100,200\}$. Also, we use \textit{Pearson Correlation} as the similarity metric for computing the similarity values between users in \algname{UserKNN} and \textit{Cosine similarity} for computing the similarity values between items in \algname{ItemKNN}. For model-based recommendation algorithms (i.e., \algname{BiasedMF}, \algname{SVD++}, \algname{ListRankMF}, and \algname{WRMF}), we set the \textit{number of factors} to $\{30,50,100,200\}$, the \textit{number of iterations} to $\{30,50,100,150,200\}$, and the \textit{learning rate} to $\{0.0001,0.001,0.01\}$. 

We use the best-performing recommendation model, to generate the recommendation list of size 10 for each user. Then, we compare the generated recommendation lists with the test data to measure the effectiveness of the recommendation model. We use \textit{Librec-Auto}~\cite{mansoury2018automating,mansoury2019algorithm} for performing our experiments and the resources are available at \url{https://github.com/masoudmansoury/Unfairness_MultifactorialBias}.

\begin{figure*}[t!]
    \centering
    \begin{subfigure}[b]{0.99\textwidth}
        \includegraphics[width=\textwidth]{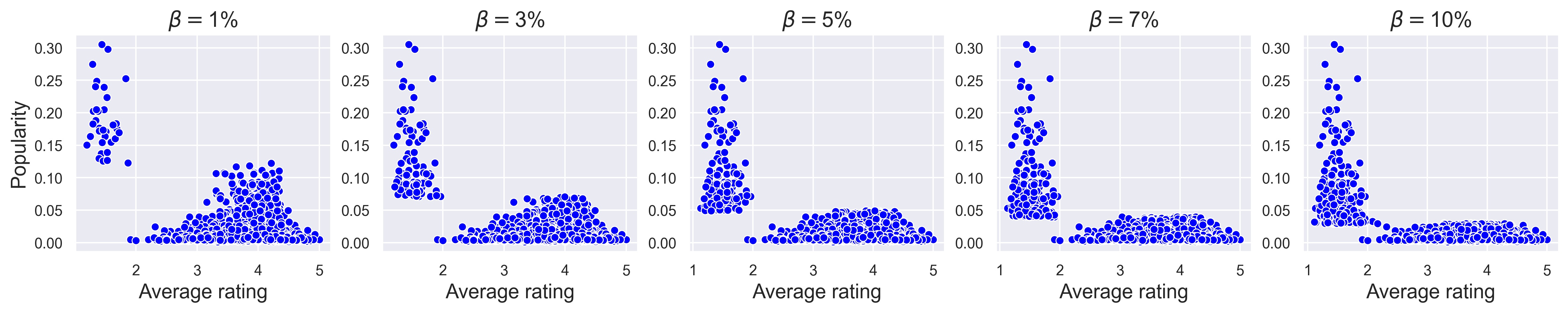}
        \caption{Goodreads}\label{fig_goodreads_pop_rating_sim}
    \end{subfigure}
    \begin{subfigure}[b]{0.99\textwidth}
        \includegraphics[width=\textwidth]{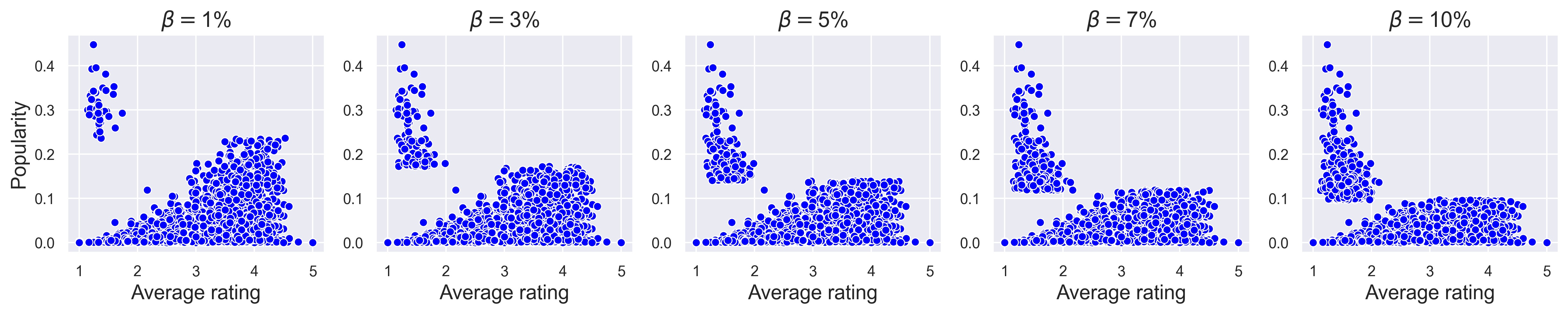}
        \caption{MovieLens}\label{fig_ml_pop_rating_sim}
    \end{subfigure}
    \begin{subfigure}[b]{0.99\textwidth}
        \includegraphics[width=\textwidth]{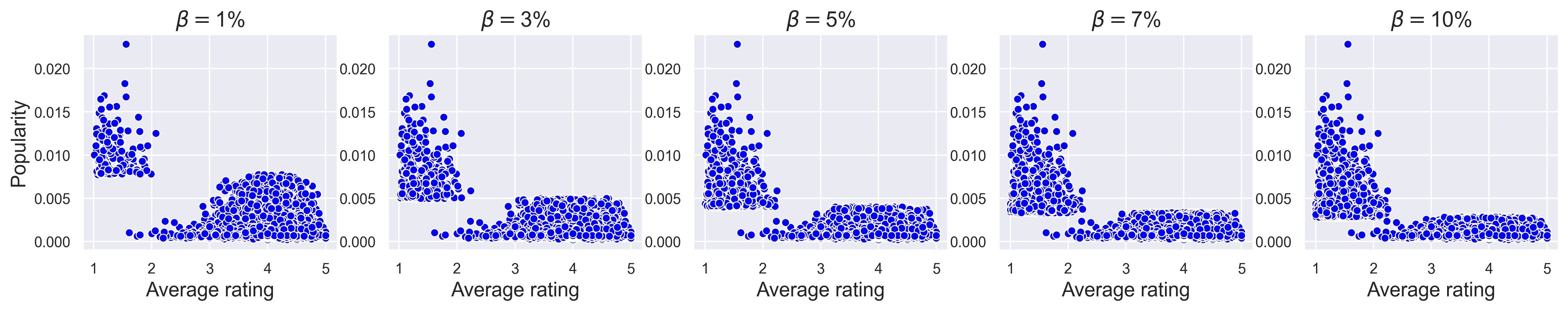}
        \caption{Google Local Data}\label{fig_glr_pop_rating_sim}
    \end{subfigure}
    \begin{subfigure}[b]{0.99\textwidth}
        \includegraphics[width=\textwidth]{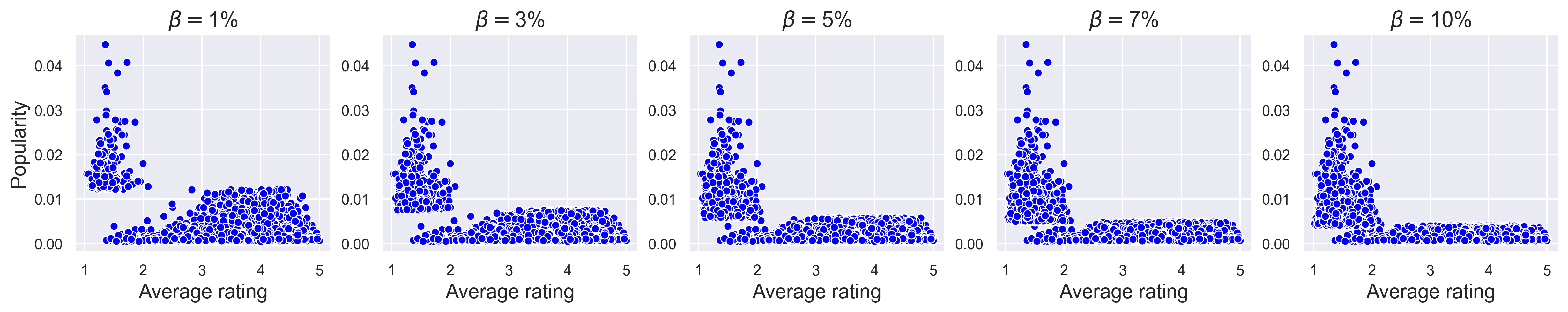}
        \caption{Yelp}\label{fig_yelp_pop_rating_sim}
    \end{subfigure}
\caption{The relationship between the average ratings and popularity of items in four datasets after artificially mitigating for popularity-positivity bias.}\label{fig_pop_rating_sim}
\end{figure*}

\section{The Impact of Multifactorial Bias in Recommendation: A Simulation Study}
\label{sec_simulation}

In this section, we conduct a simulation study on the impact of multifactorial bias on recommendations. In this simulation, we artificially change the high rating values assigned to most popular items to low rating values in the input rating data as a pre-processing step and then compare the performance of recommendation models on the original and modified data. For example, on the MovieLens dataset, we modify the rating data by changing rating 5 (i.e., the highest rating) given to popular items to 1 (i.e., the lowest rating) before feeding it into the recommendation model.

What makes this simulation interesting is that the modified data has the same characteristics as the original data (e.g., popularity bias) and only differs in ratings given to the popular items. This signifies that for the modified data, we synthetically mitigated the multifactorial bias. Therefore, the degree of multifactorial bias is the major difference between the modified data and the original data which we use to show the impact of multifactorial bias on recommendations.

To reliably observe the impact of multifactorial bias, we repeat the simulation by modifying the ratings given to $\beta$\% of most popular items with varying values of $\beta$, i.e., $\beta \in \{1\%, 3\%, 5\%, 7\%, 10\%, 15\%, 20\%, 25\%, 30\%, 35\%, 40\%, 45\%, 50\%\}$. Figure \ref{fig_pop_rating_sim} shows the relationship between the average ratings and popularity of each item on four datasets after the aforementioned pre-processing for different $\beta$ values (i.e., $\beta \in \{1\%, 3\%, 5\%, 7\%, 10\%\}$). As expected, with increasing $\beta$ value, more high ratings given to the popular items are converted to low rating values. Note that this conversion does not change the shape of the popularity distribution in each of these datasets, and the only difference is on the positivity bias on popular items (i.e., multifactorial bias). 

\begin{figure*}[t!]
    \centering
    \begin{subfigure}[b]{0.99\textwidth}
        \includegraphics[width=\textwidth]{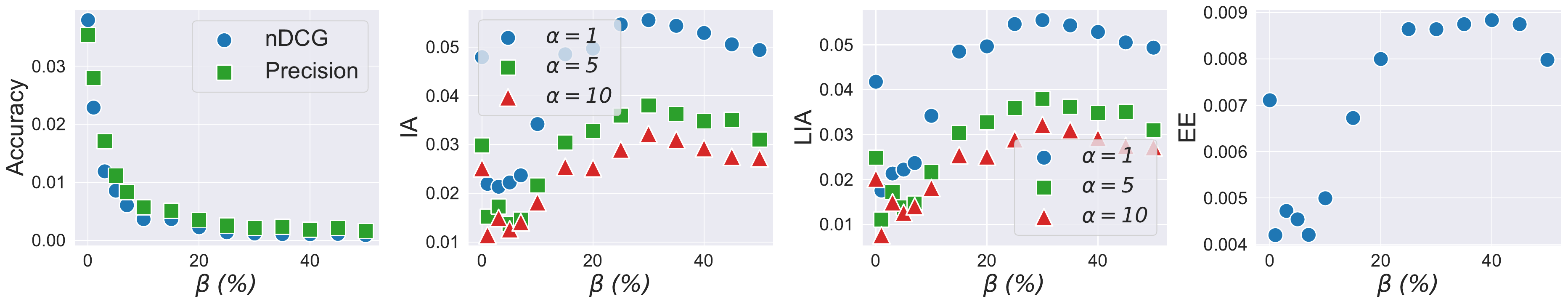}
        \caption{Goodreads}\label{fig_goodreads_biasedmf_sim}
    \end{subfigure}
    \begin{subfigure}[b]{0.99\textwidth}
        \includegraphics[width=\textwidth]{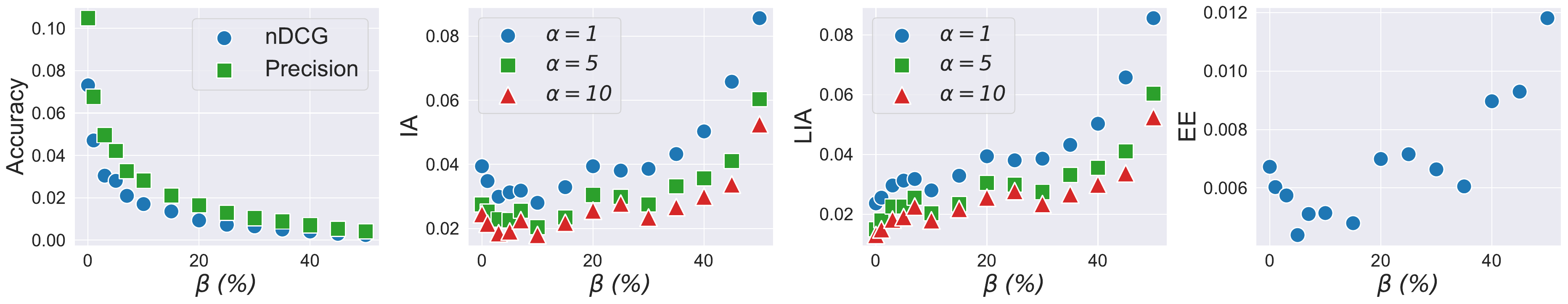}
        \caption{MovieLens}\label{fig_ml_biasedmf_sim}
    \end{subfigure}
    \begin{subfigure}[b]{0.99\textwidth}
        \includegraphics[width=\textwidth]{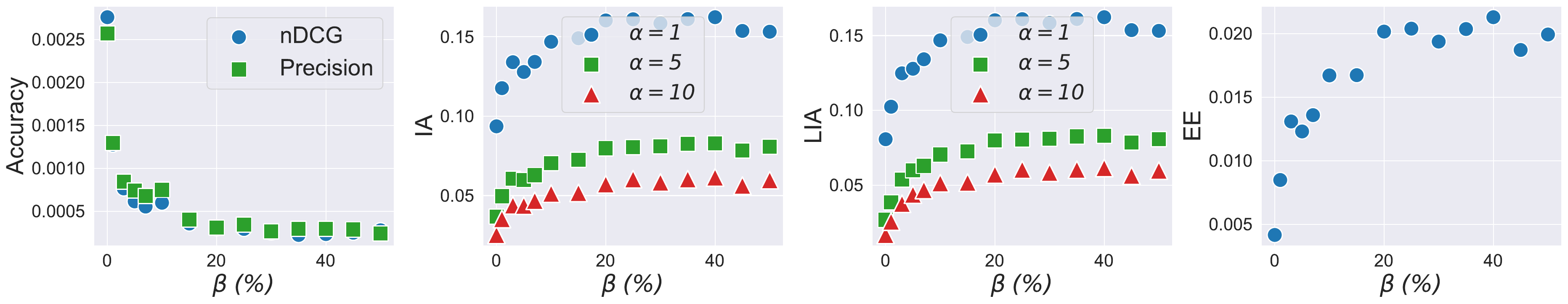}
        \caption{Google Local Data}\label{fig_glr_biasedmf_sim}
    \end{subfigure}
    \begin{subfigure}[b]{0.99\textwidth}
        \includegraphics[width=\textwidth]{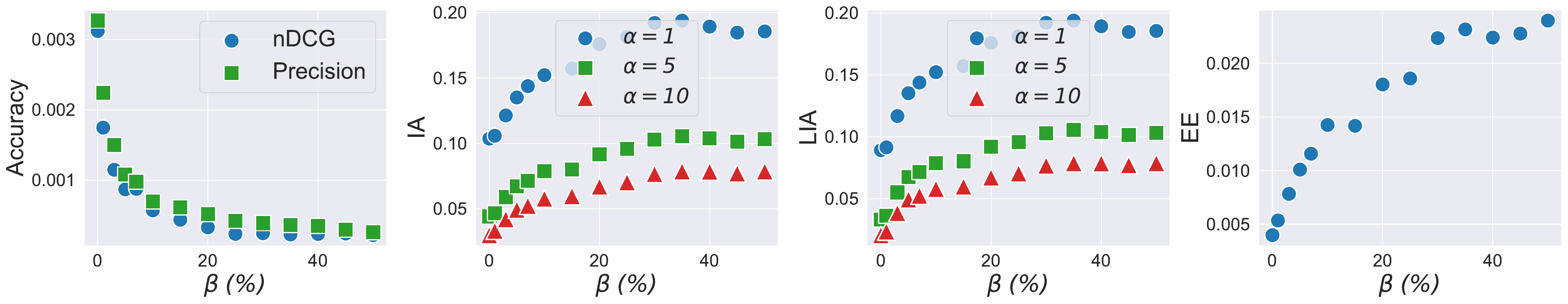}
        \caption{Yelp}\label{fig_yelp_biasedmf_sim}
    \end{subfigure}
\caption{The performance of recommendation model on modified rating data in our simulation study on three datasets. Experiments are performed using \algname{BiasedMF} algorithm.}\label{fig_biasedmf_sim}
\end{figure*}

Figure~\ref{fig_biasedmf_sim} shows the results of our simulation in terms of both accuracy and bias metrics. In all plots, the x-axis indicates the $\beta$ value, percentage of most popular items whose high ratings are changed to low ratings, and the y-axis shows the metric value, either accuracy (e.g., precision and nDCG) or fairness metrics (e.g., IA, LIA, and EE).

According to the plots in Figure~\ref{fig_biasedmf_sim}, as $\beta$ increases, both precision and nDCG are decreasing, possibly because less popular items are getting recommended by this change in the ratings of popular items. This loss in  accuracy of the recommender system may not necessarily implies a negative consequence as the test set also suffers popularity bias. On the other hand, from the plots showing item aggregate diversity (IA), long-tail aggregate diversity (LIA), and equality of exposure (EE), it is evident that the fairness of recommendations is improved. These results suggest the effect of positivity bias on popular items, reducing the accuracy of the recommendations (due to evaluation with a biased test set) while improving the fairness of exposure for items in recommendation results.
\section{Mitigating Unfairness of Multifactorial Bias in Recommendation}
\label{sec_percentile}

As shown in Section~\ref{sec_simulation}, while mitigating multifactorial bias helps to improve the exposure fairness of the items in recommendation results, it negatively affects the recommendation accuracy. This indicates that blindly mitigating multifactorial bias may not properly address the problem in real-world scenarios and that a principled solution is needed to address the trade-off between exposure fairness and accuracy of recommendation results.

We base our bias mitigation approach on the rating transformation method proposed by \citet{mansoury2021flatter}. The authors introduce a rating transformation method that converts the ratings in a user's profile into percentile values to reduce the positivity bias in users' rating behavior. They show that the converted rating data results in more uniform distribution for ratings in users' profiles and consequently leads to improved ranking quality for the recommendation results. In this paper, we also build on the same transformation method to reduce multifactorial bias in rating data. Unlike \citet{mansoury2021flatter}, who perform a percentile transformation on user profiles, we perform a percentile transformation on the profiles of items to mitigate the positivity bias, particularly on popular items. 

Given $R_{.,i}$ as the ratings given by different users on item $i$, the percentile value corresponding to each rating in $R_{.,i}$ is calculated as:
\begin{equation}\label{eq_percentile}
    \mathit{Per}(r, R_{.,i})=\frac{100 \times \mathit{position}(r,o(R_{.,i}))}{|R_{.,i}|+1}, \quad\forall r \in R_{.,i},
\end{equation}
\noindent where $\mathit{Per}(r, R_{.,i})$ is the percentile value corresponding to rating $r$ in $R_{.,i}$, $o(R_{.,i})$ is sorted $R_{.,i}$ in the ascending order, and $\mathit{position}(r,o(R_{.,i}))$ returns the index of $r$ in $o(R_{.,i})$. One challenge in this transformation is that $R_{.,i}$ often contains repeated ratings which makes specifying the position of a specific rating unclear. For example, in $R_{.,i}=\left<1,2,2,2,2,3,3,4,5,5\right>$, it is not clear what the position of rating 2 should be. Possible solutions for this issue can be returning the first occurrence (e.g., position 2), last occurrence (e.g., position 5), or something in between. Consistent with the findings in \cite{mansoury2021flatter}, our experiments show that considering the last occurrence as the position of the repeated rating yields the best result. Therefore, for the remainder of the paper, our percentile transformation is based on the last occurrence of the repeated ratings.

On each dataset used in this paper (described in Section~\ref{sec_dataset}), we transform the rating data into percentile data using the proposed percentile transformation method. In other words, for each user $u$ and item $i$ in rating matrix $R$, we transform the rating $R_{ui}$ into the percentile value using Eq. \ref{eq_percentile}. This will create a user-item percentile matrix. We denote this matrix as $P \in \R^{n \times m}$ with $P_{ui}$ being the percentile value corresponding to $R_{ui}$. 

\begin{figure*}[t!]
    \centering
    \begin{subfigure}[b]{0.42\textwidth}
        \includegraphics[width=\textwidth]{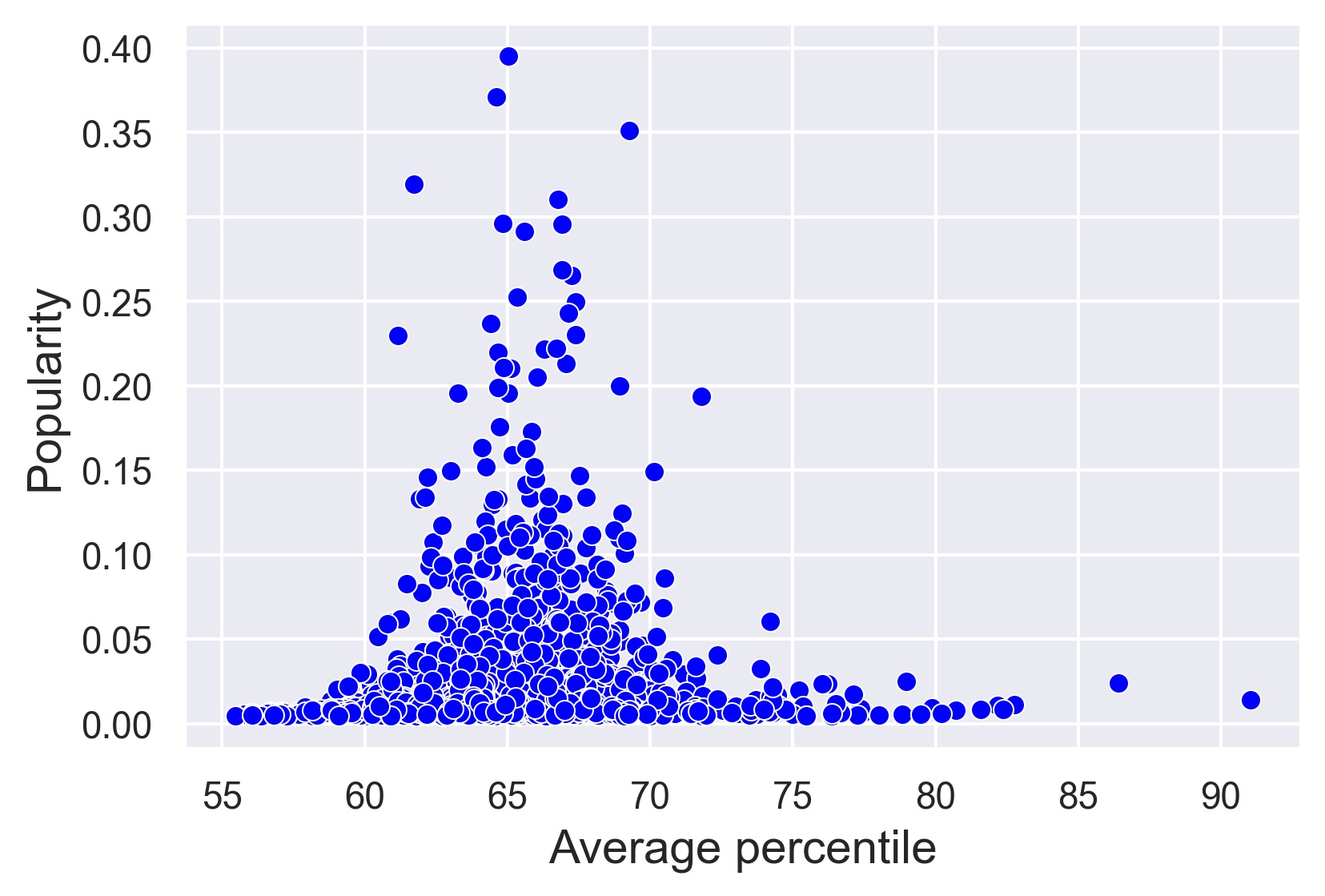}
        \caption{Goodreads}\label{fig_goodreads_pop_percentile}
    \end{subfigure}
    \begin{subfigure}[b]{0.42\textwidth}
        \includegraphics[width=\textwidth]{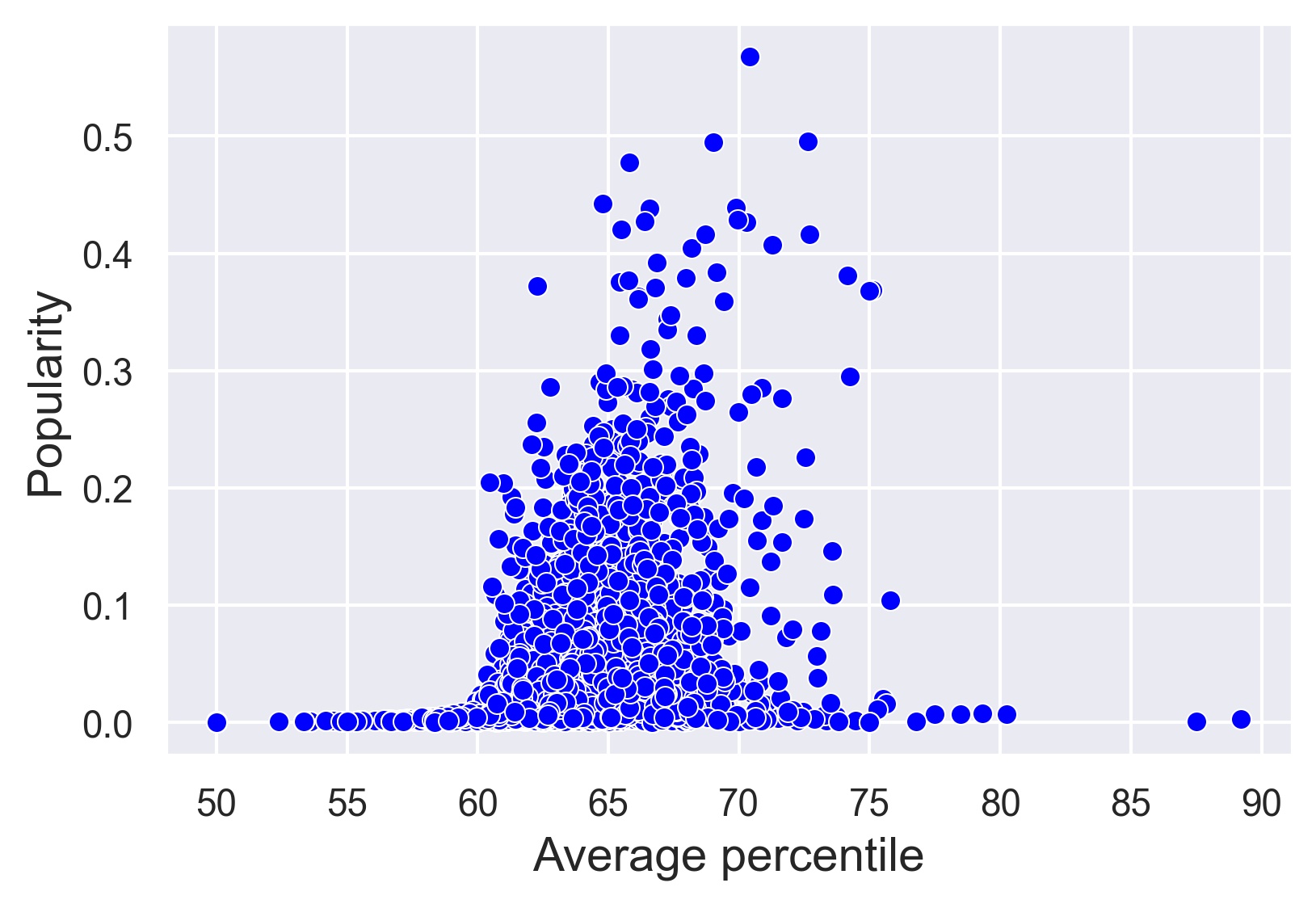}
        \caption{MovieLens}\label{fig_ml_pop_percentile}
    \end{subfigure}
    \begin{subfigure}[b]{0.42\textwidth}
        \includegraphics[width=\textwidth]{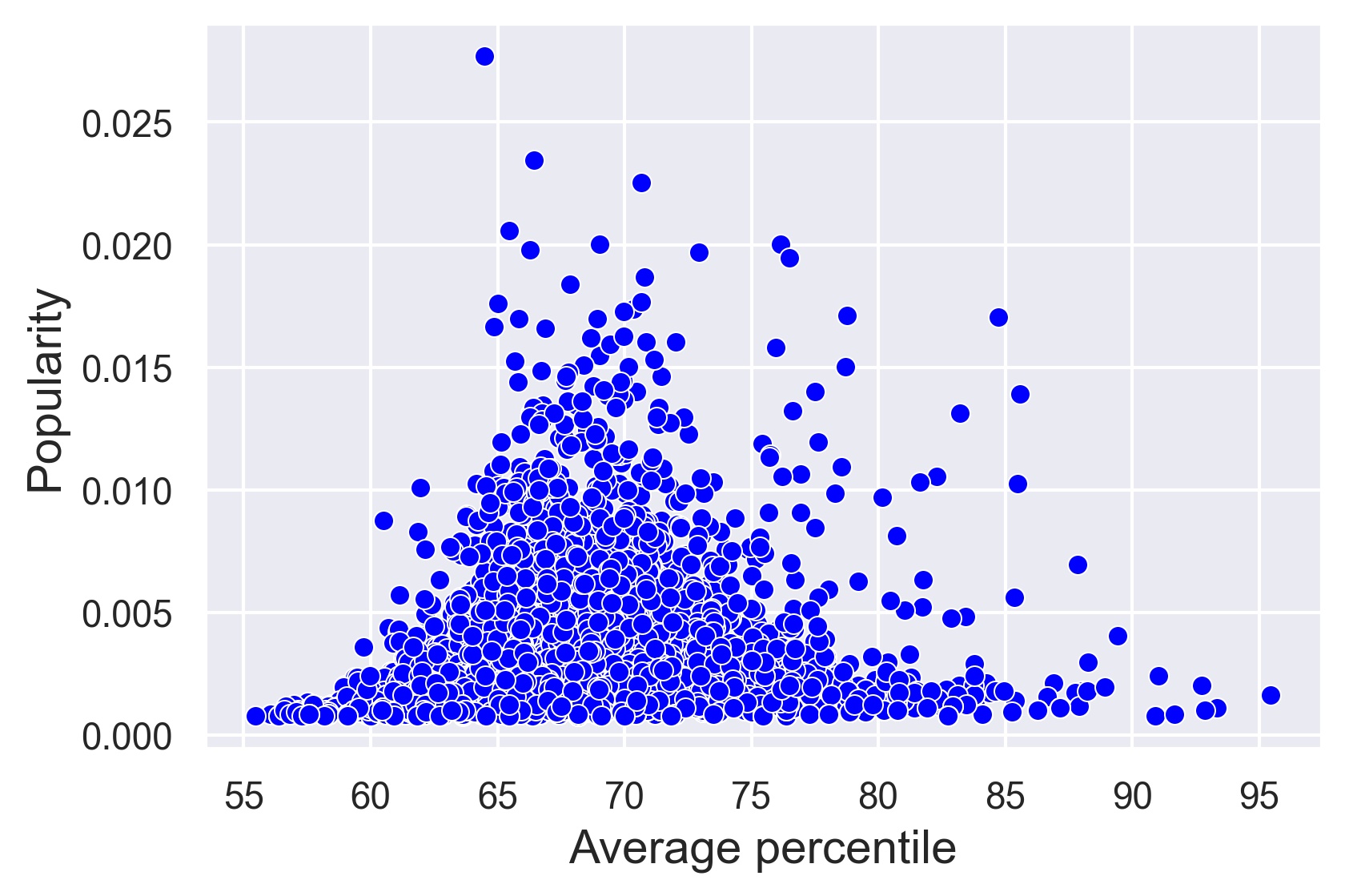}
        \caption{Google Local Data}\label{fig_glr_pop_percentile}
    \end{subfigure}
    \begin{subfigure}[b]{0.42\textwidth}
        \includegraphics[width=\textwidth]{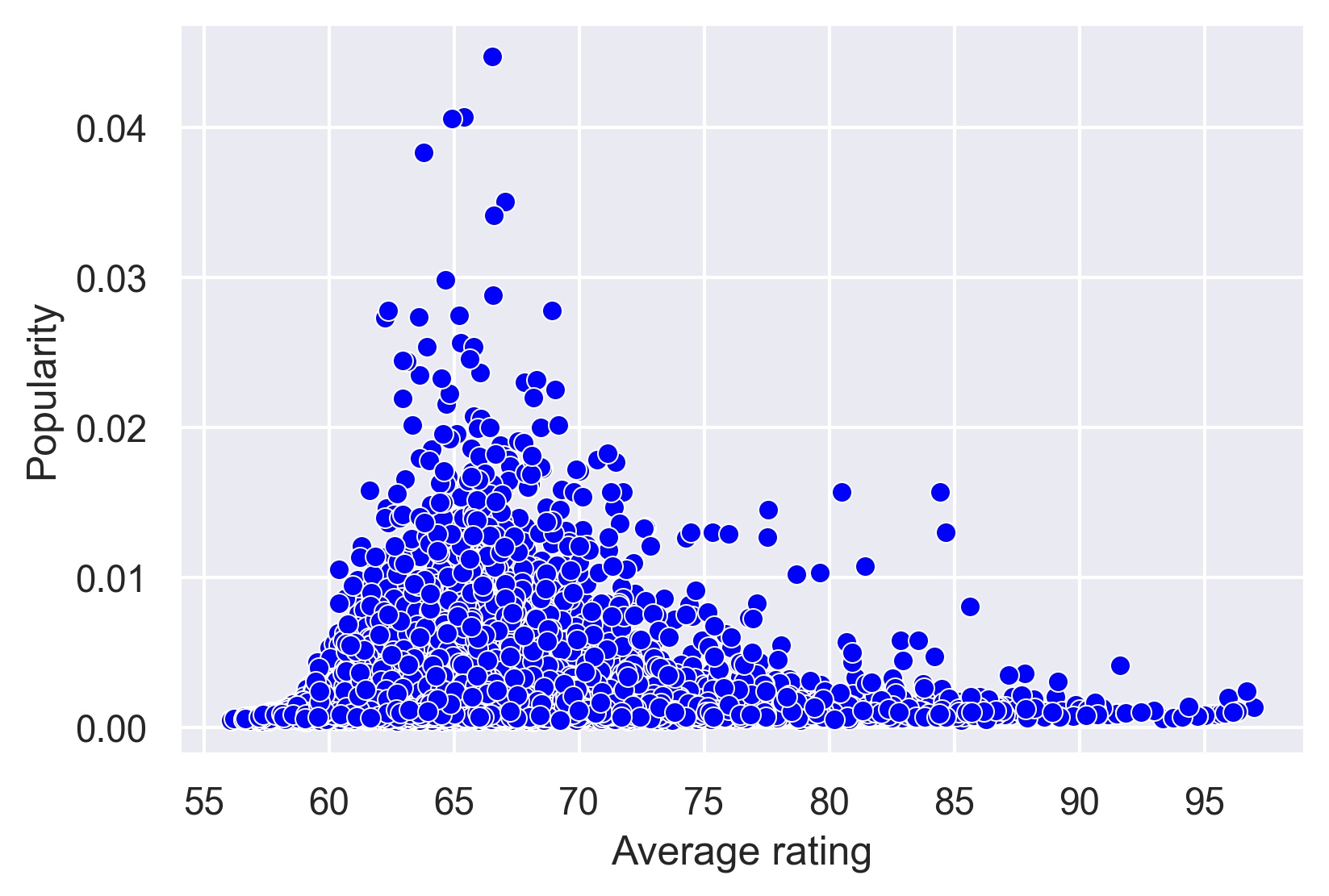}
        \caption{Yelp}\label{fig_yelp_pop_percentile}
    \end{subfigure}
\caption{The relationship between the average percentile values and popularity of each item. }\label{fig_pop_percentile}
\end{figure*}

To see to what degree the transformed percentile data suffers multifactorial bias, we re-produce the plots in Figure~\ref{fig_pop_rating} using the percentile matrix. Figure~\ref{fig_pop_percentile} shows the relationship between the popularity and the average percentile value of items on four datasets. As shown, on all datasets, the percentile matrix does not show the patterns of multifactorial bias: the average percentile values for both popular and unpopular items are concentrated on certain values and there is no situation where high percentile values are being assigned to a few popular items, as was the case in rating data in Figure~\ref{fig_pop_rating}. This means that in the percentile matrix, the rating values are more equally represented.

\begin{figure*}[t!]
    \centering
    \begin{subfigure}[c]{1\textwidth}
        \centering
        \includegraphics[width=.3\textwidth]{figures/per_legend.pdf}
        \vspace{5pt}
    \end{subfigure}
    \\
    \begin{subfigure}[b]{0.45\textwidth}
        \includegraphics[width=\textwidth]{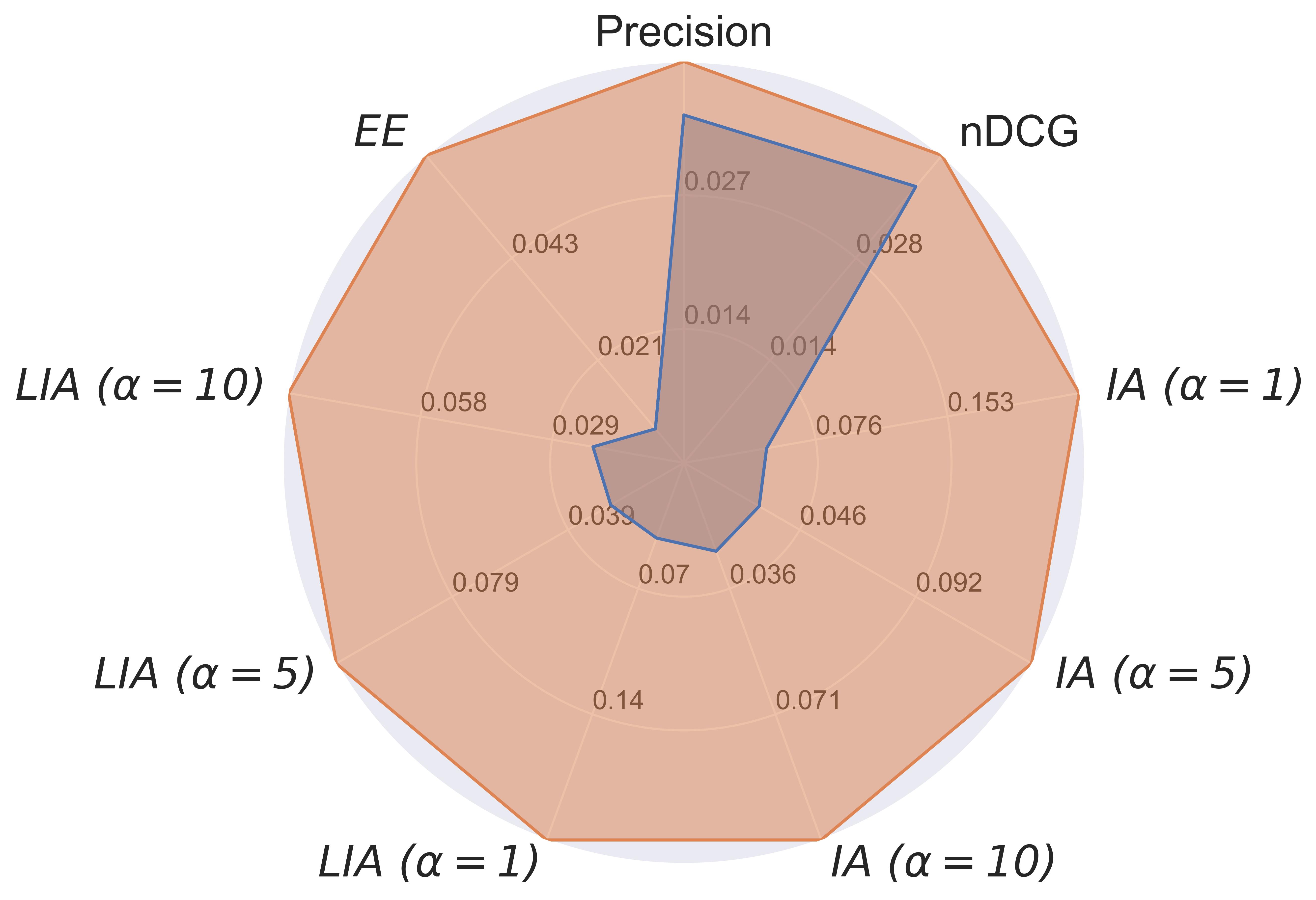}
        \caption{BiasedMF}
    \end{subfigure}
    \qquad
    \begin{subfigure}[b]{0.45\textwidth}
        \includegraphics[width=\textwidth]{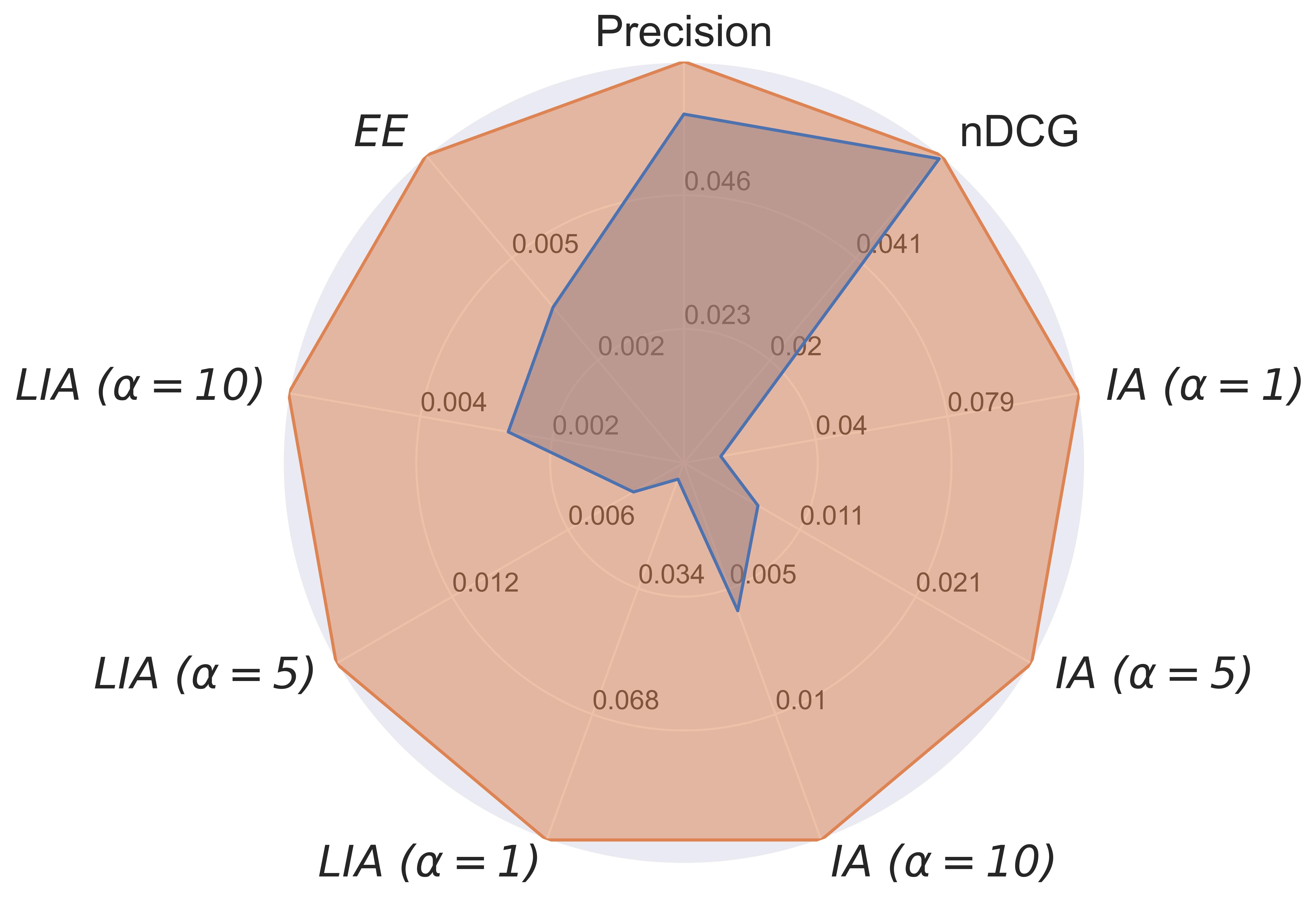}
        \caption{ListRankMF}
    \end{subfigure}
    \\
    \begin{subfigure}[b]{0.45\textwidth}
        \includegraphics[width=\textwidth]{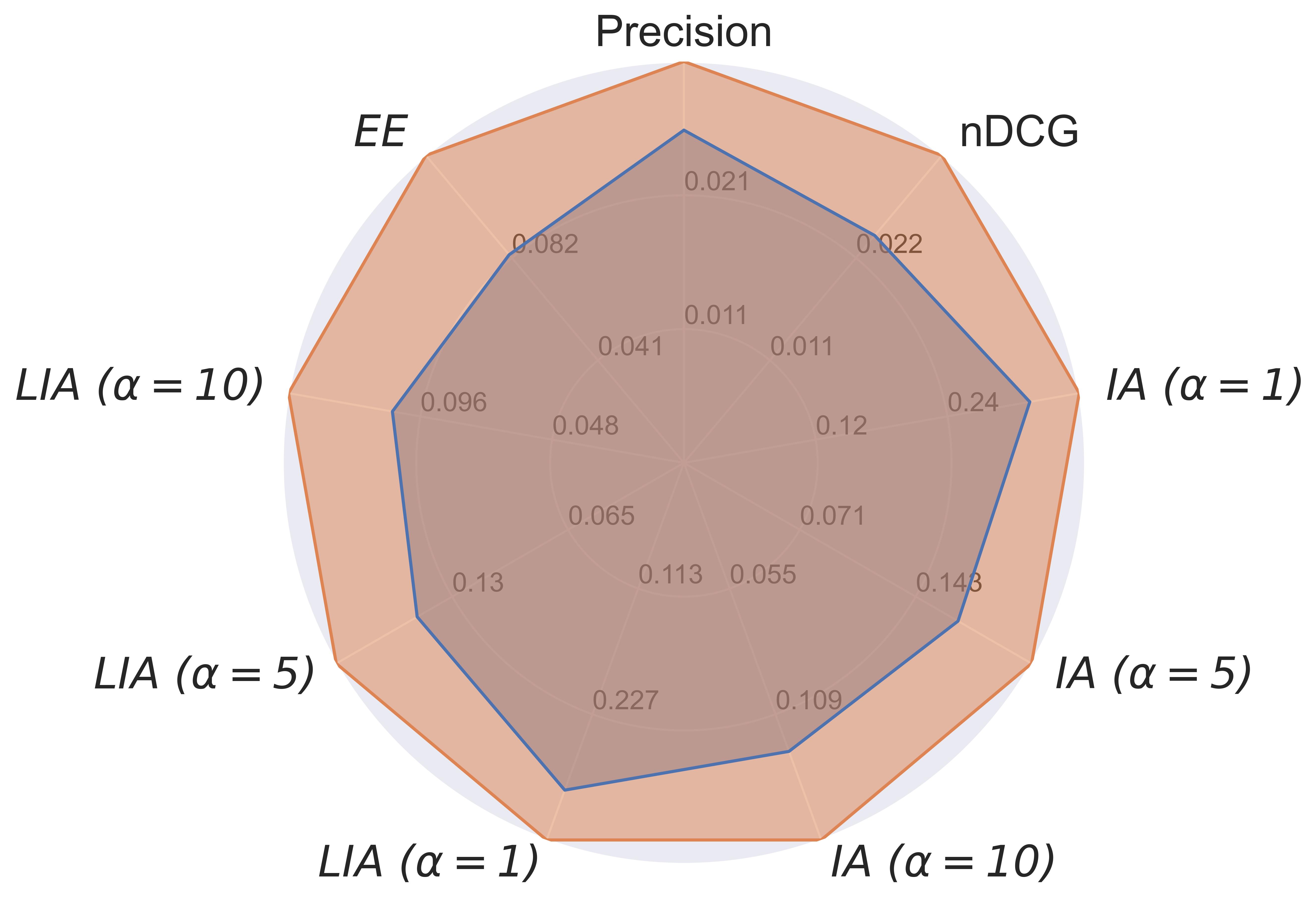}
        \caption{SVD++}
    \end{subfigure}
    \qquad
    \begin{subfigure}[b]{0.45\textwidth}
        \includegraphics[width=\textwidth]{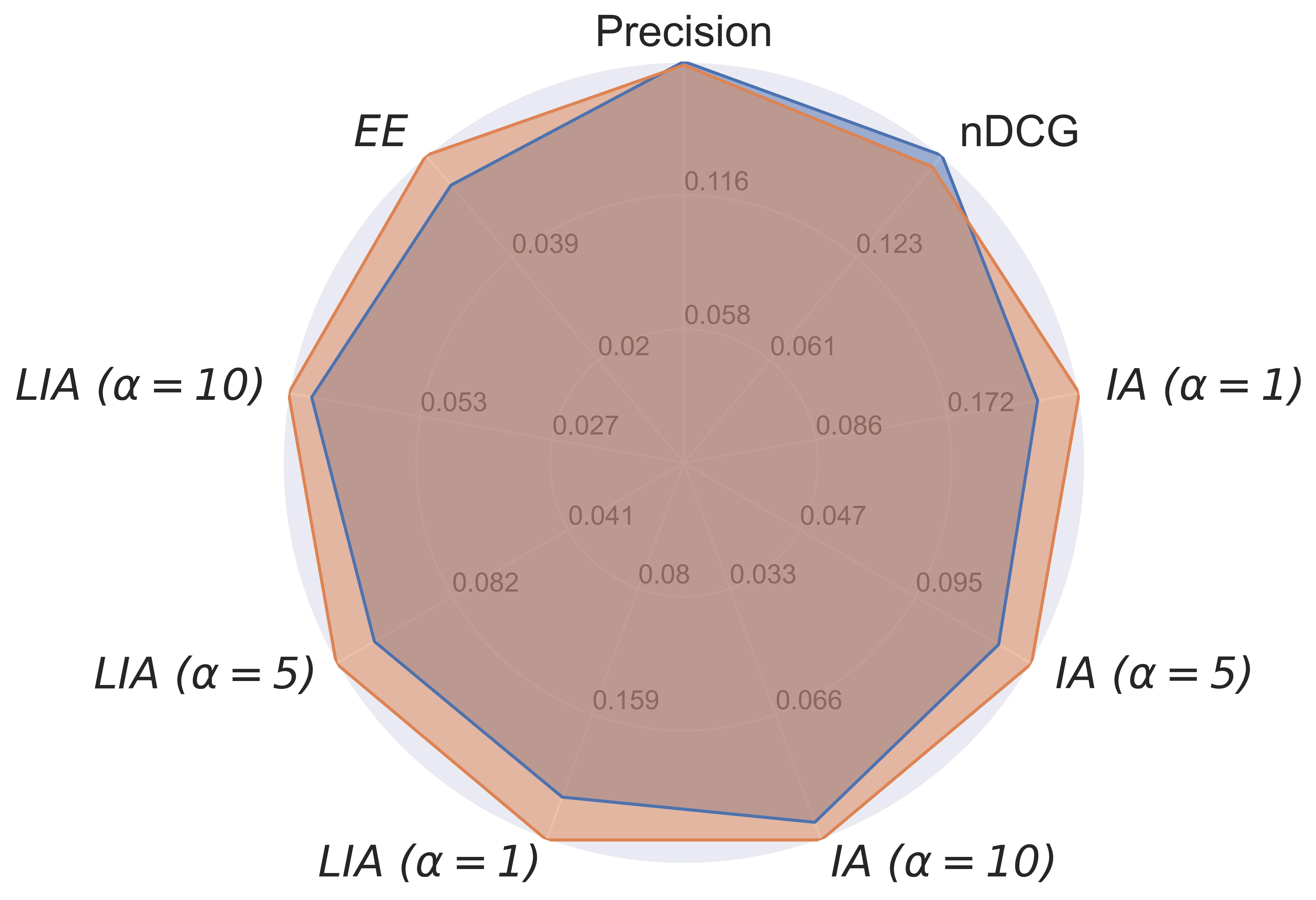}
        \caption{WRMF}
    \end{subfigure}
    \\
    \begin{subfigure}[b]{0.45\textwidth}
        \includegraphics[width=\textwidth]{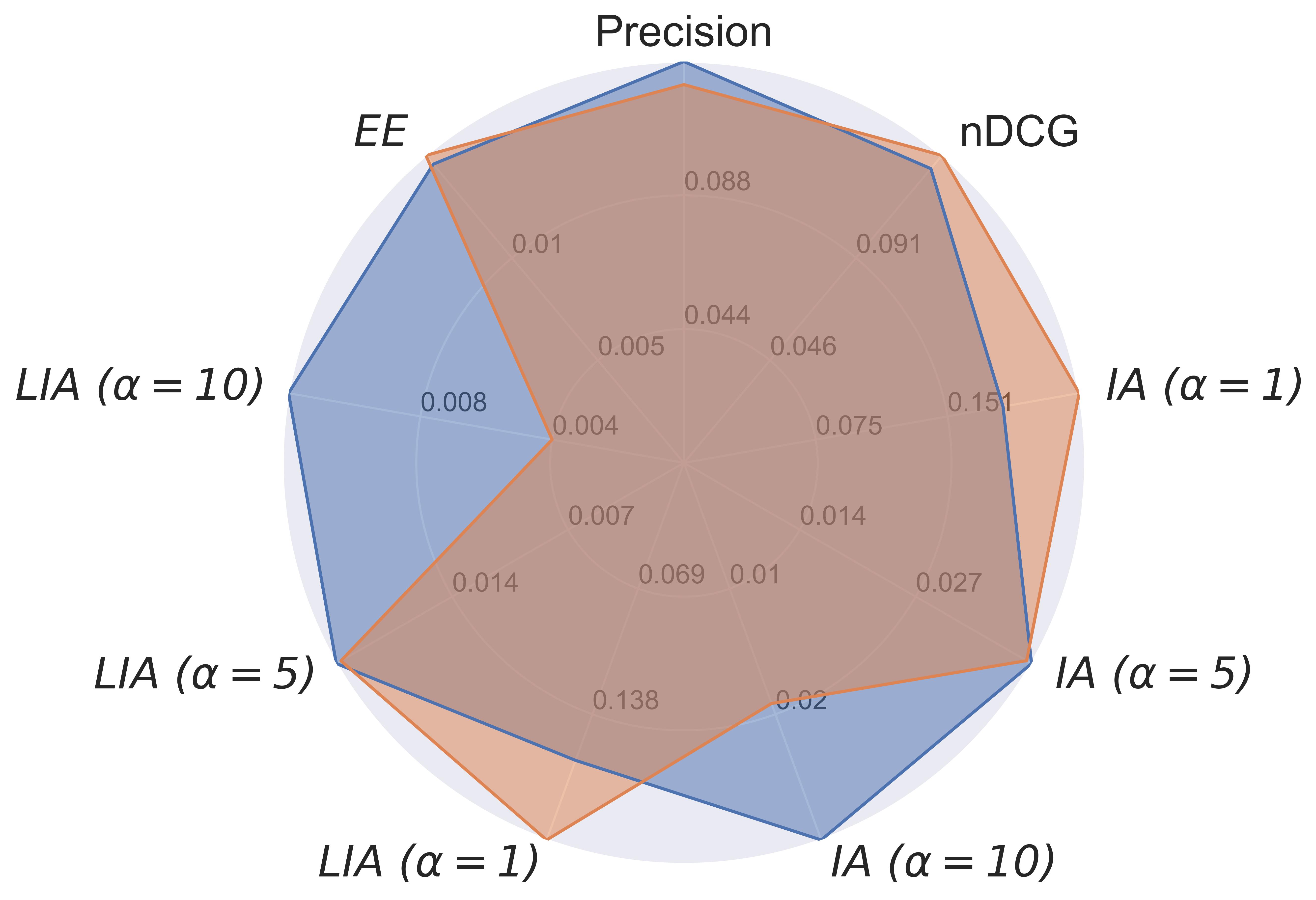}
        \caption{UserKNN}
    \end{subfigure}
    \qquad
    \begin{subfigure}[b]{0.45\textwidth}
        \includegraphics[width=\textwidth]{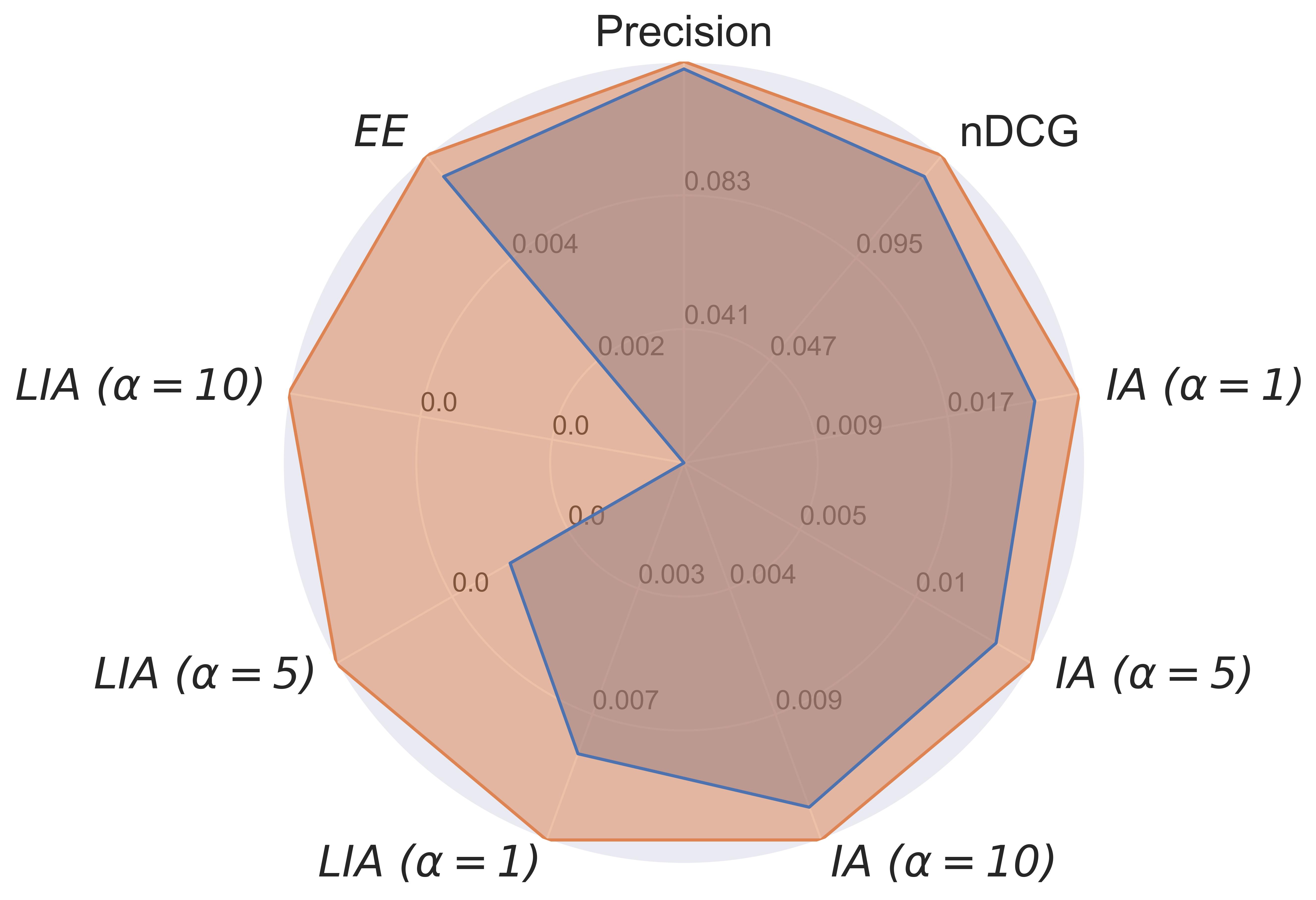}
        \caption{ItemKNN}
    \end{subfigure}
\caption{Performance of six recommender algorithms when raw rating data and percentile data are separately used as input on the Goodreads dataset. For all metrics, higher values signify a better performance.}\label{fig_goodreads_pop_rating_sim}
\end{figure*}

\begin{figure*}[t!]
    \centering
    \begin{subfigure}[c]{1\textwidth}
        \centering
        \includegraphics[width=.3\textwidth]{figures/per_legend.pdf}
        \vspace{5pt}
    \end{subfigure}
    \\
    \begin{subfigure}[b]{0.45\textwidth}
        \includegraphics[width=\textwidth]{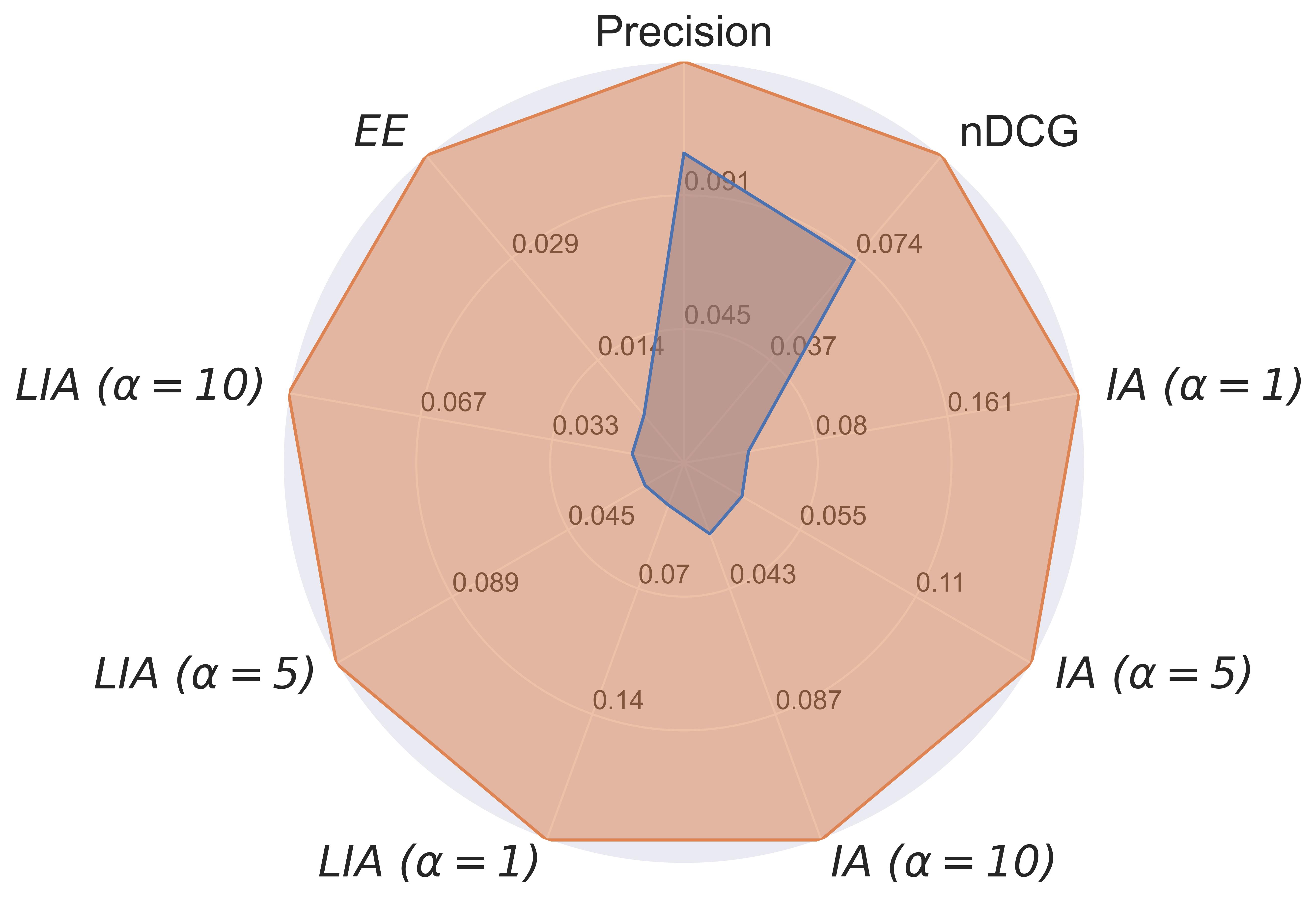}
        \caption{BiasedMF}
    \end{subfigure}
    \qquad
    \begin{subfigure}[b]{0.45\textwidth}
        \includegraphics[width=\textwidth]{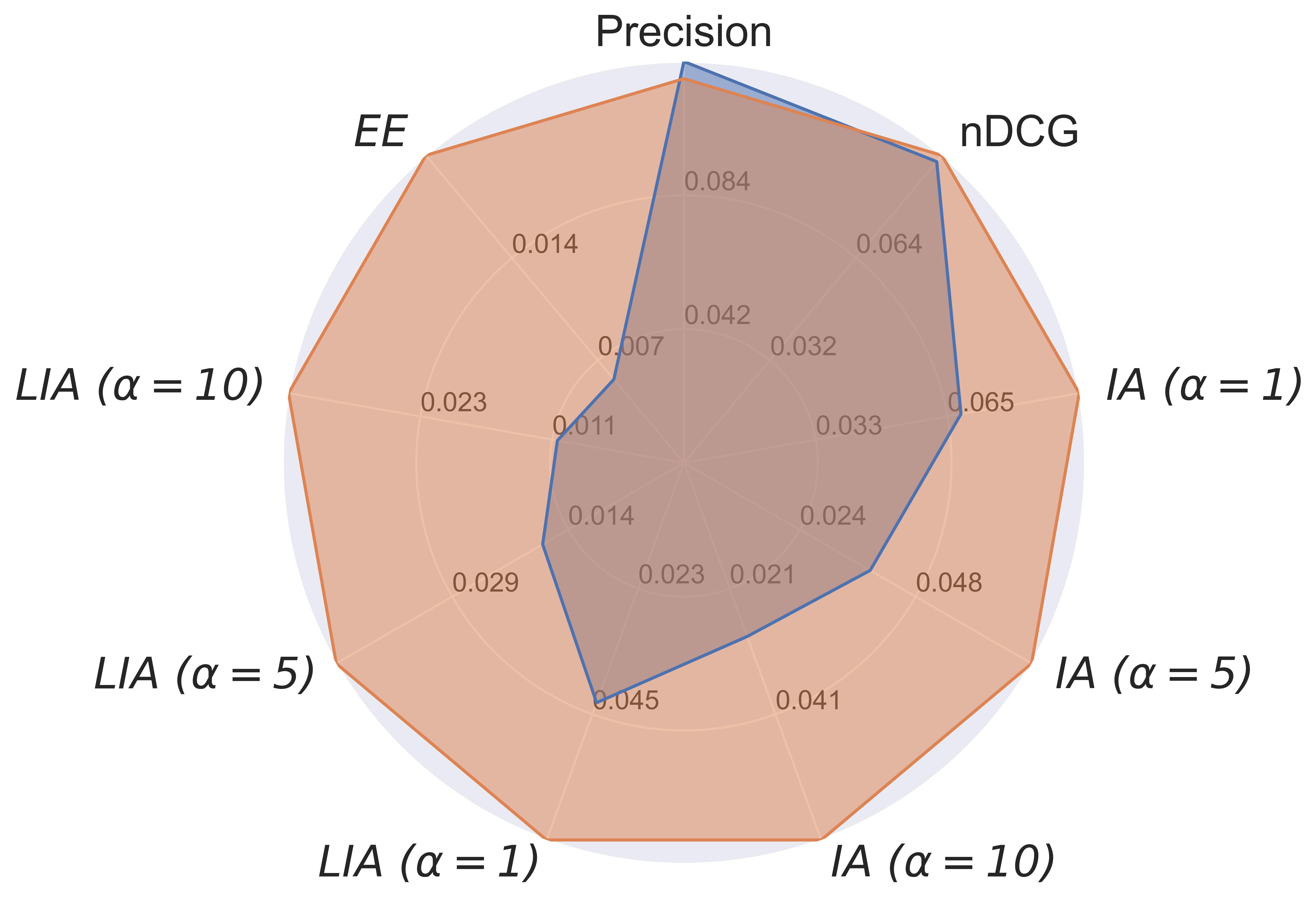}
        \caption{ListRankMF}
    \end{subfigure}
    \\
    \begin{subfigure}[b]{0.45\textwidth}
        \includegraphics[width=\textwidth]{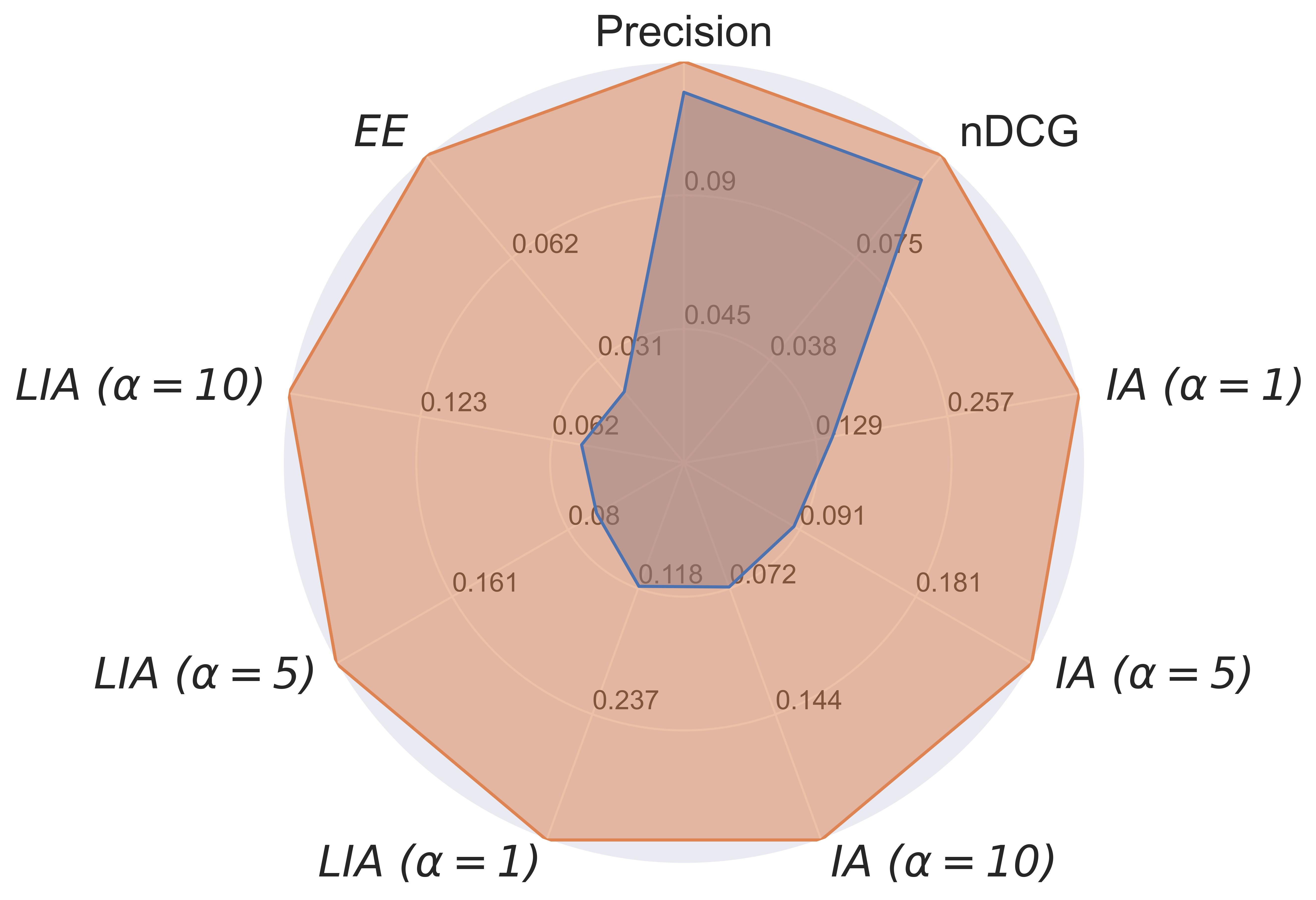}
        \caption{SVD++}
    \end{subfigure}
    \qquad
    \begin{subfigure}[b]{0.45\textwidth}
        \includegraphics[width=\textwidth]{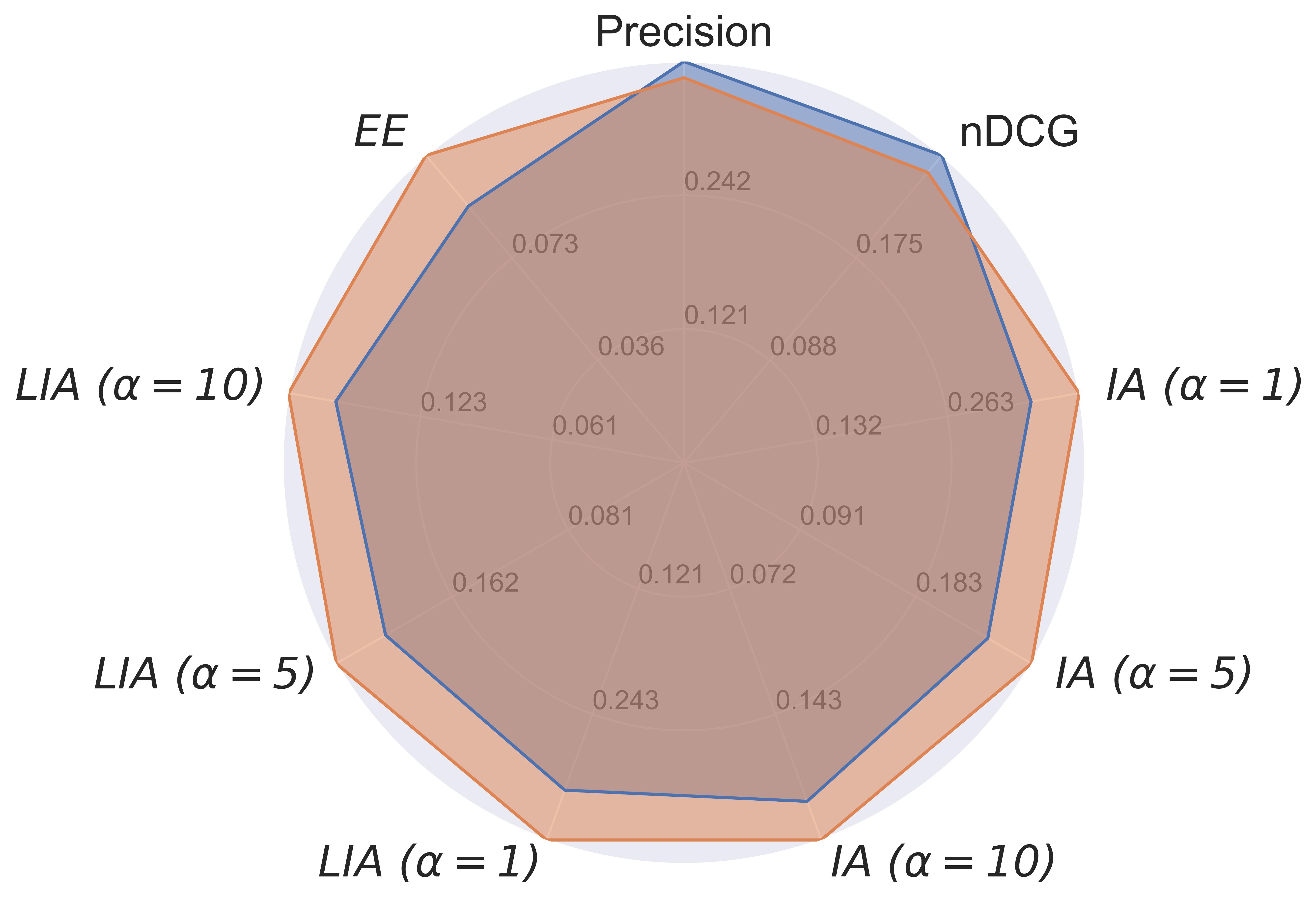}
        \caption{WRMF}
    \end{subfigure}
    \\
    \begin{subfigure}[b]{0.45\textwidth}
        \includegraphics[width=\textwidth]{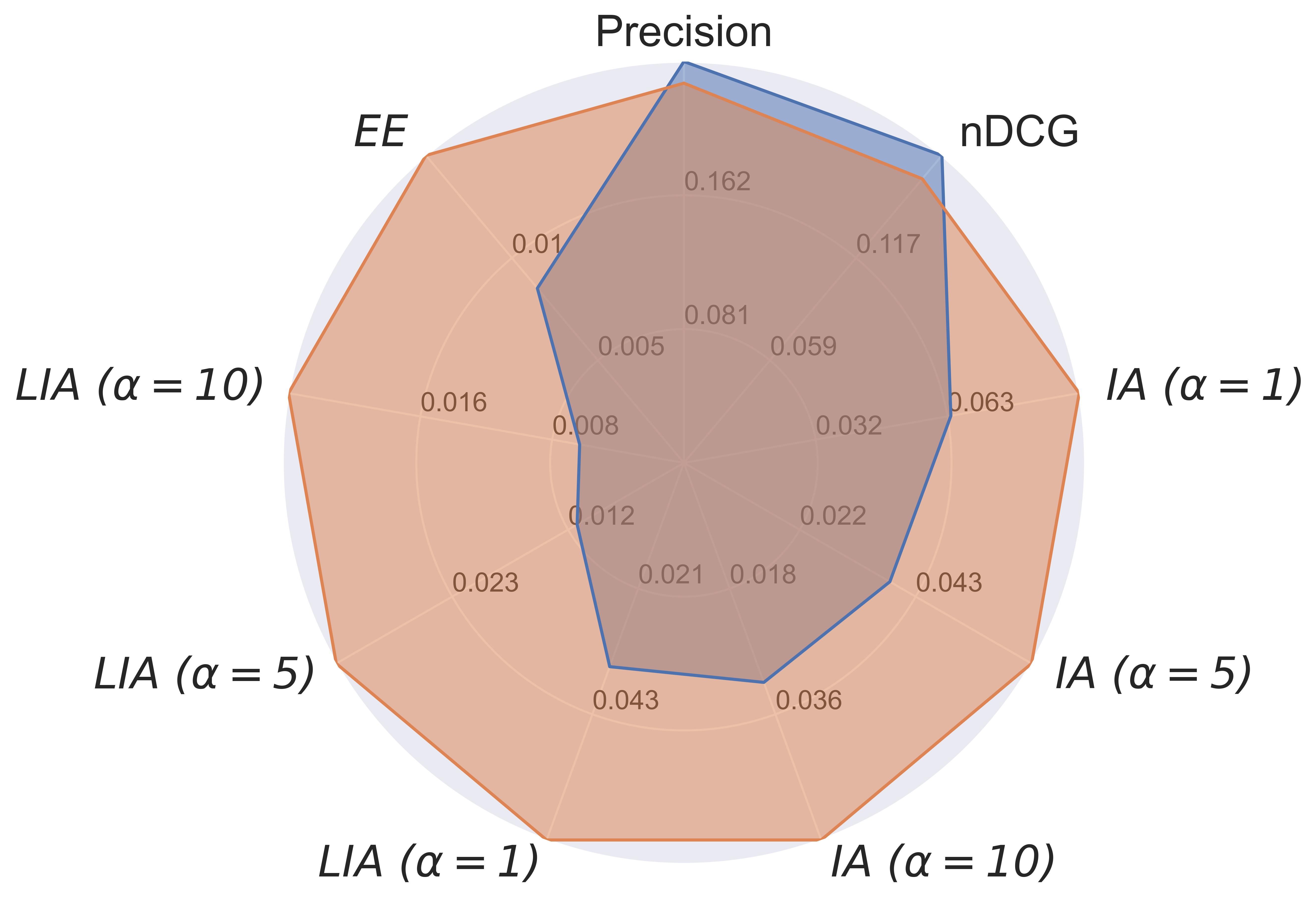}
        \caption{UserKNN}
    \end{subfigure}
    \qquad
    \begin{subfigure}[b]{0.45\textwidth}
        \includegraphics[width=\textwidth]{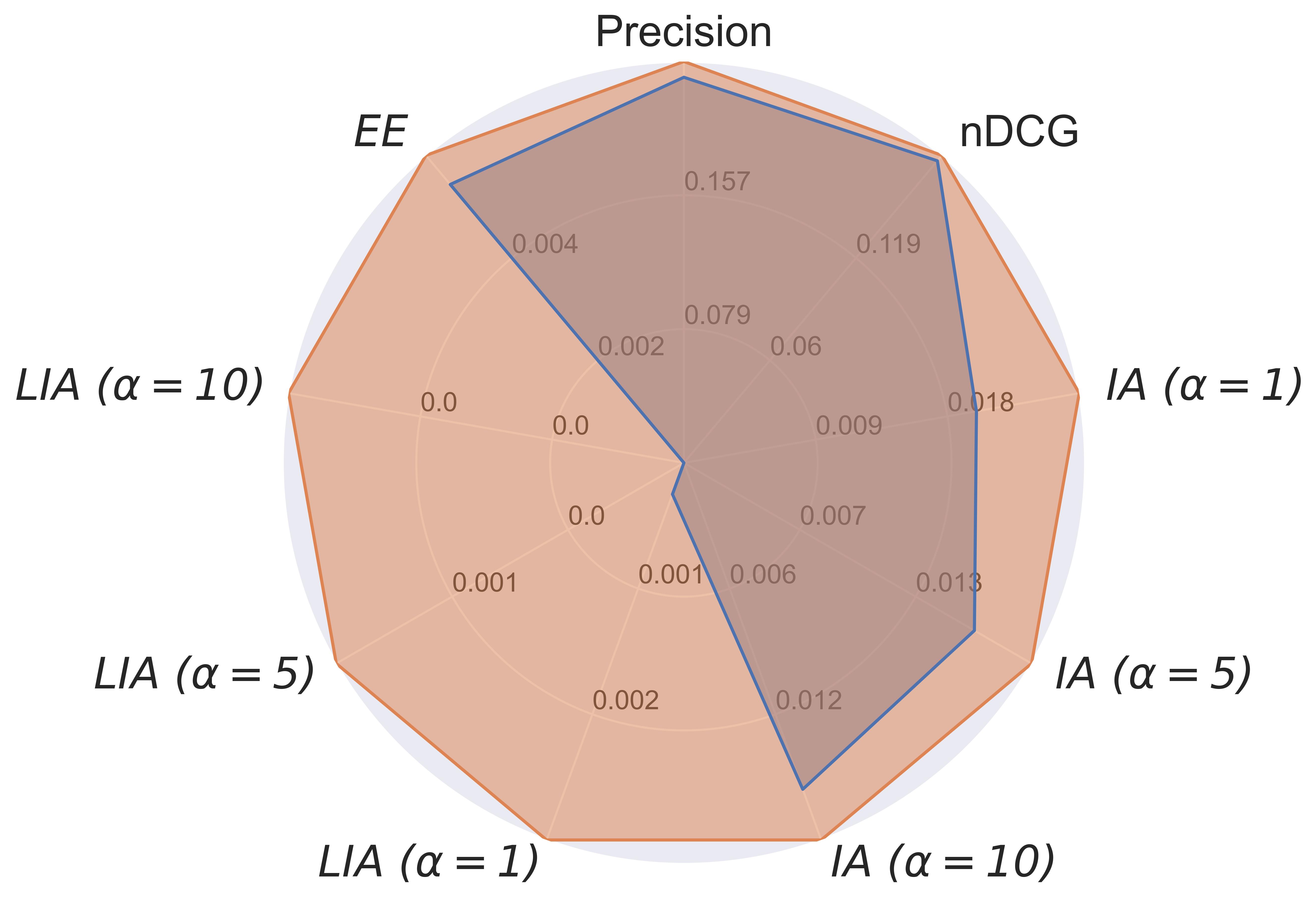}
        \caption{ItemKNN}
    \end{subfigure}
\caption{Performance of six recommendation algorithms when raw rating data and percentile data are separately used as input on the MovieLens dataset. For all metrics, higher values signify a better performance.}\label{fig_ml_pop_rating_sim}
\end{figure*}

\begin{figure*}[t!]
    \centering
    \begin{subfigure}[c]{1\textwidth}
        \centering
        \includegraphics[width=.3\textwidth]{figures/per_legend.pdf}
        \vspace{5pt}
    \end{subfigure}
    \\
    \begin{subfigure}[b]{0.45\textwidth}
        \includegraphics[width=\textwidth]{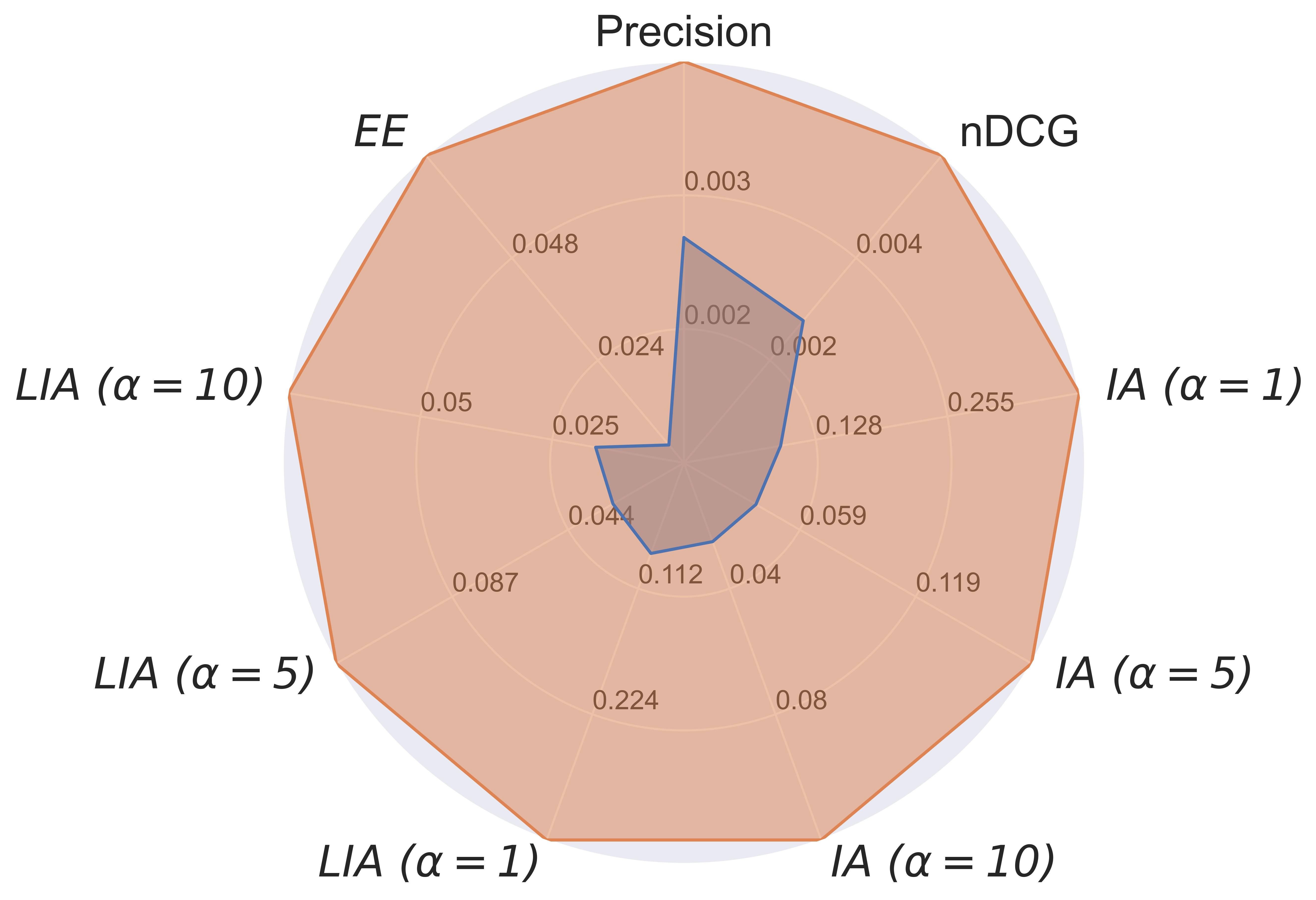}
        \caption{BiasedMF}
    \end{subfigure}
    \qquad
    \begin{subfigure}[b]{0.45\textwidth}
        \includegraphics[width=\textwidth]{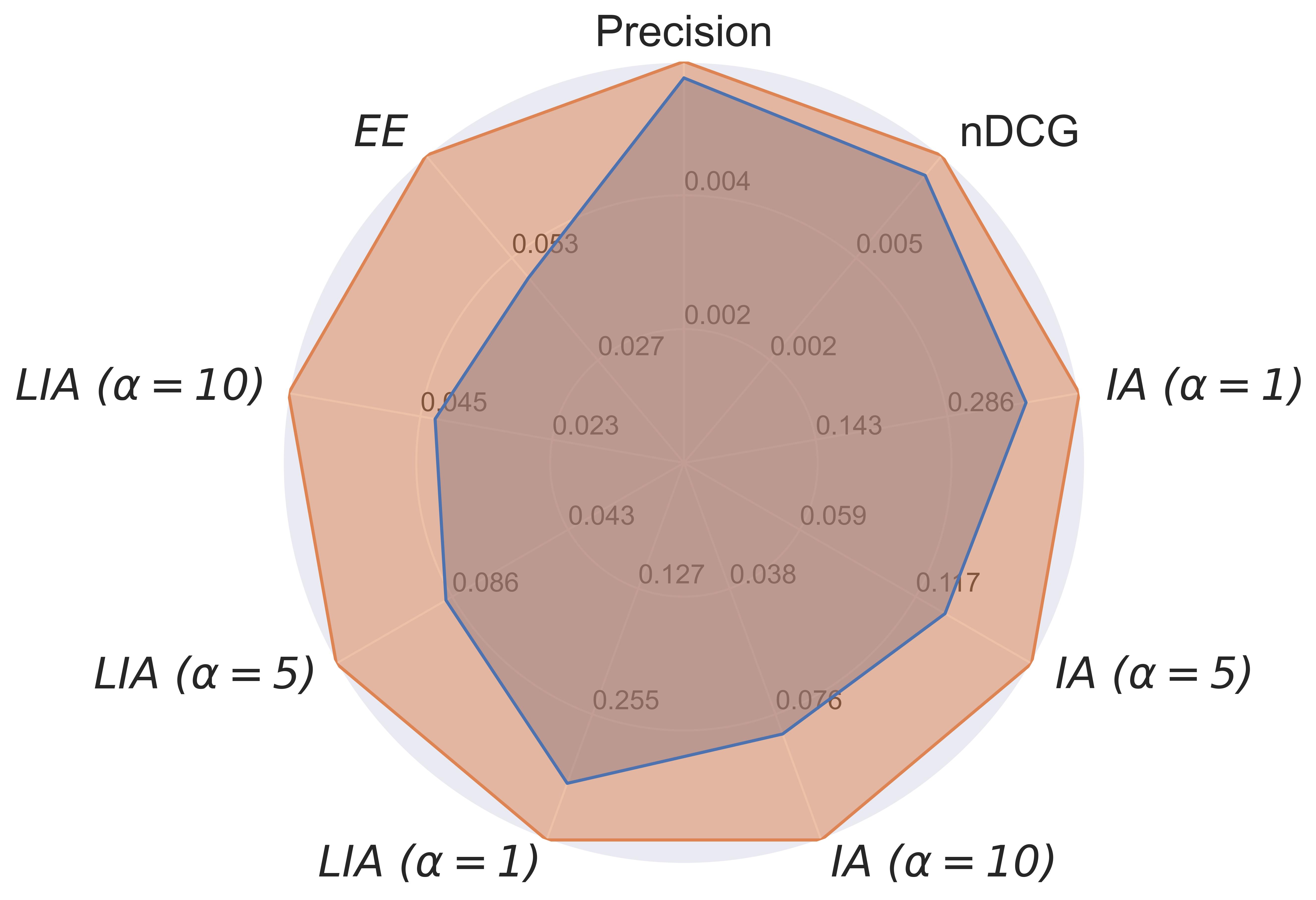}
        \caption{ListRankMF}
    \end{subfigure}
    \\
    \begin{subfigure}[b]{0.45\textwidth}
        \includegraphics[width=\textwidth]{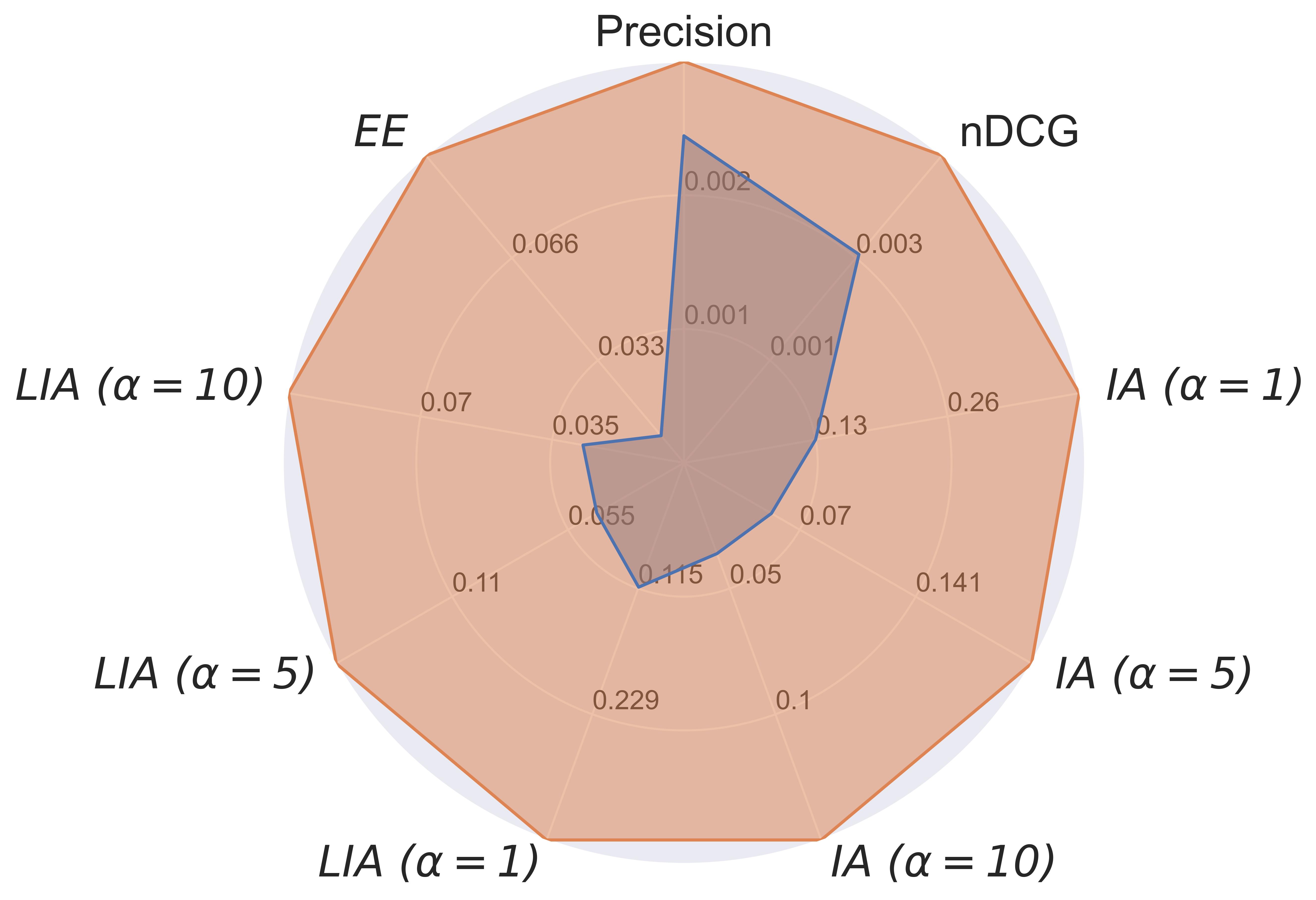}
        \caption{SVD++}
    \end{subfigure}
    \qquad
    \begin{subfigure}[b]{0.45\textwidth}
        \includegraphics[width=\textwidth]{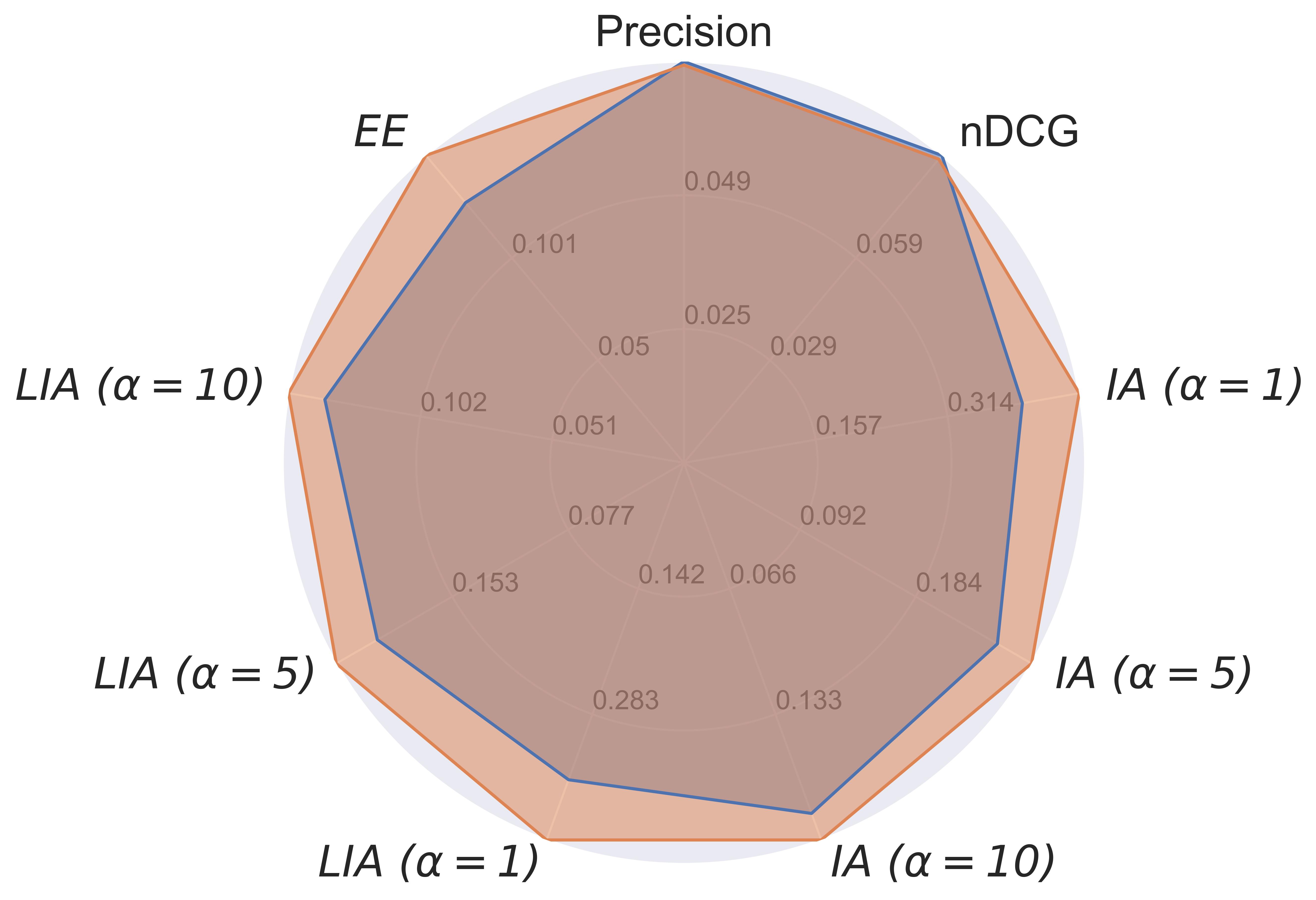}
        \caption{WRMF}
    \end{subfigure}
    \\
    \begin{subfigure}[b]{0.45\textwidth}
        \includegraphics[width=\textwidth]{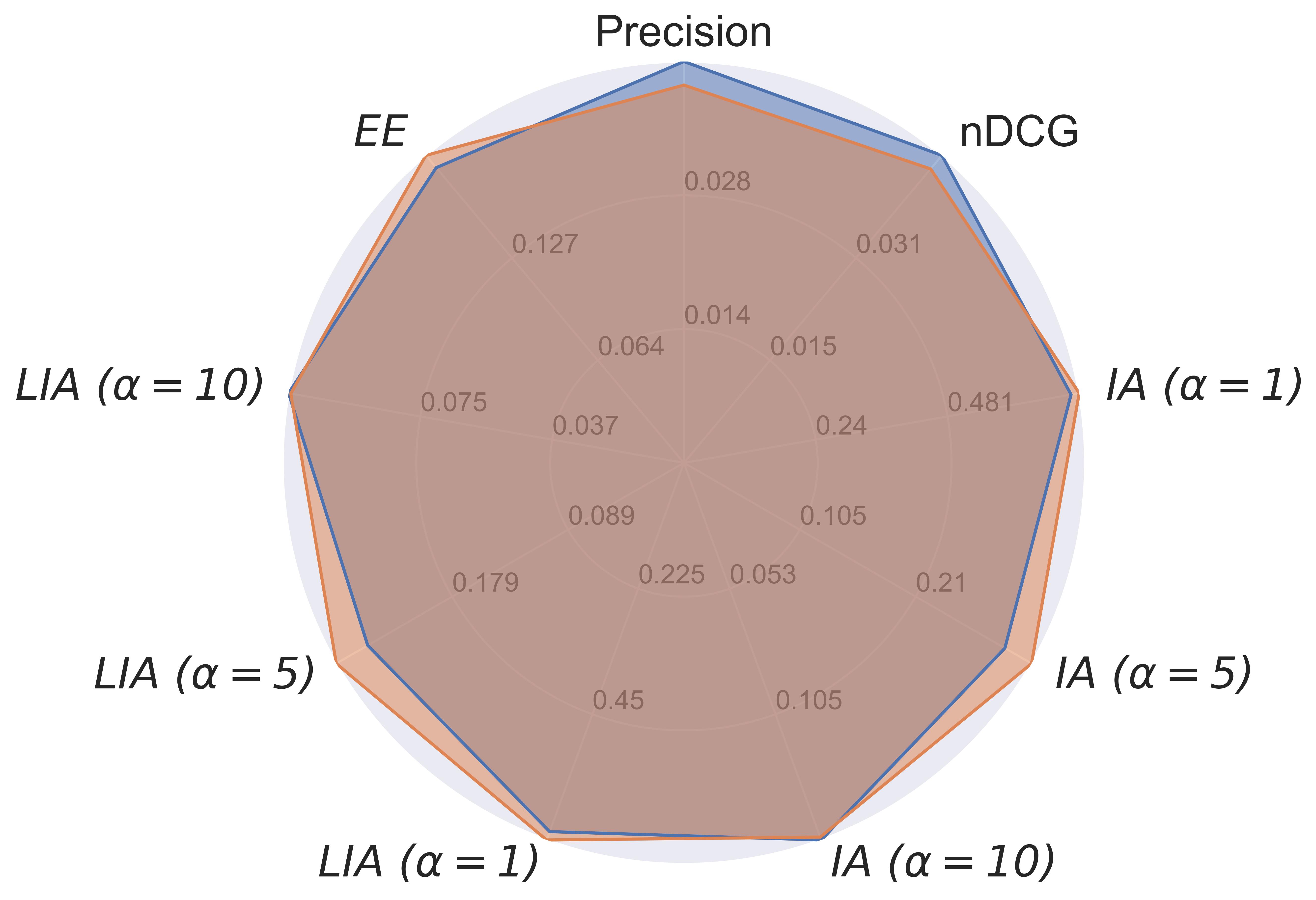}
        \caption{UserKNN}
    \end{subfigure}
    \qquad
    \begin{subfigure}[b]{0.45\textwidth}
        \includegraphics[width=\textwidth]{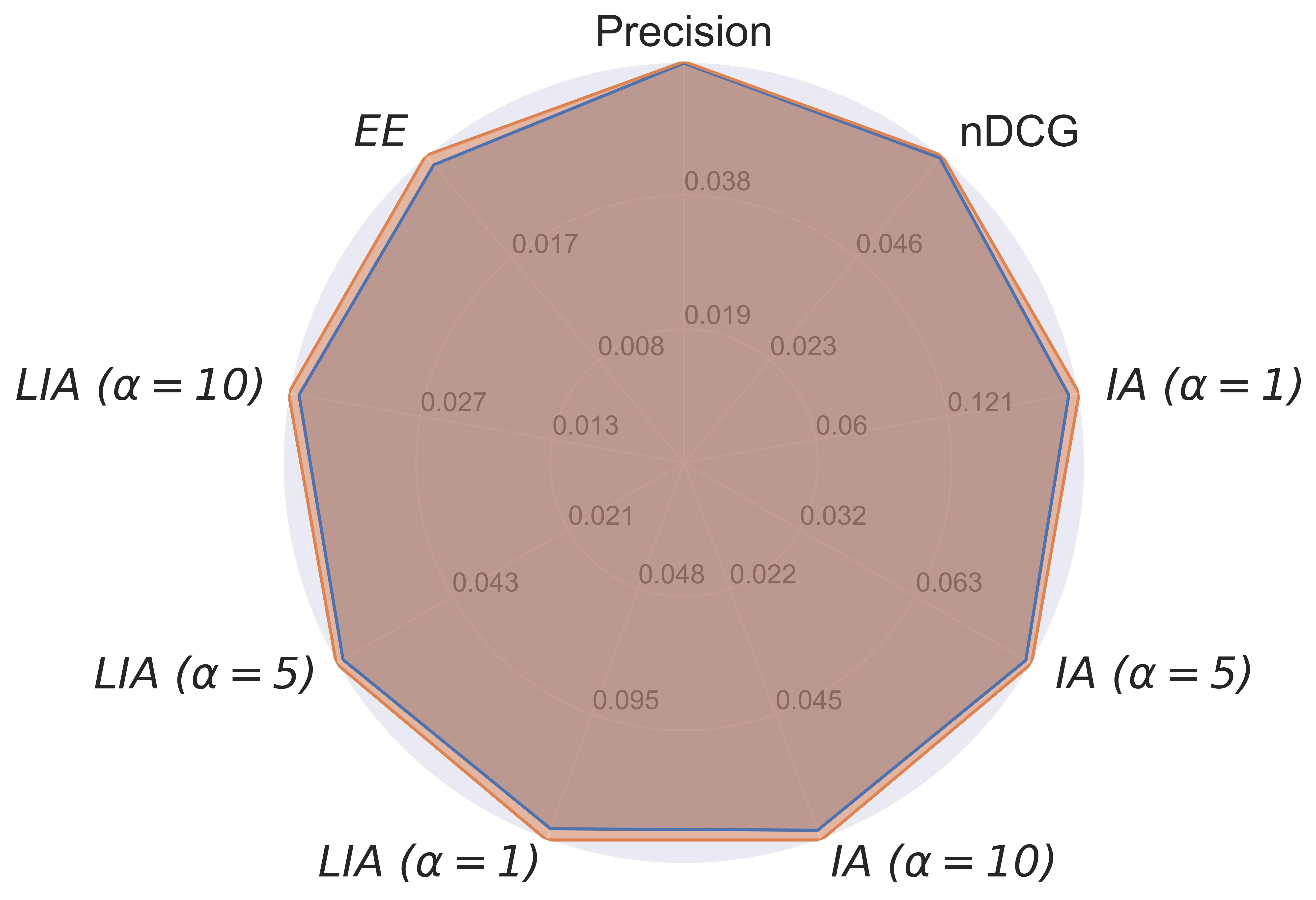}
        \caption{ItemKNN}
    \end{subfigure}
\caption{Performance of six recommendation algorithms when raw rating data and percentile data are separately used as input on the Google Local Data dataset. For all metrics, higher values signify a better performance.}\label{fig_glr_pop_rating_sim}
\end{figure*}

\begin{figure*}[t!]
    \centering
    \begin{subfigure}[c]{1\textwidth}
        \centering
        \includegraphics[width=.3\textwidth]{figures/per_legend.pdf}
        \vspace{5pt}
    \end{subfigure}
    \\
    \begin{subfigure}[b]{0.45\textwidth}
        \includegraphics[width=\textwidth]{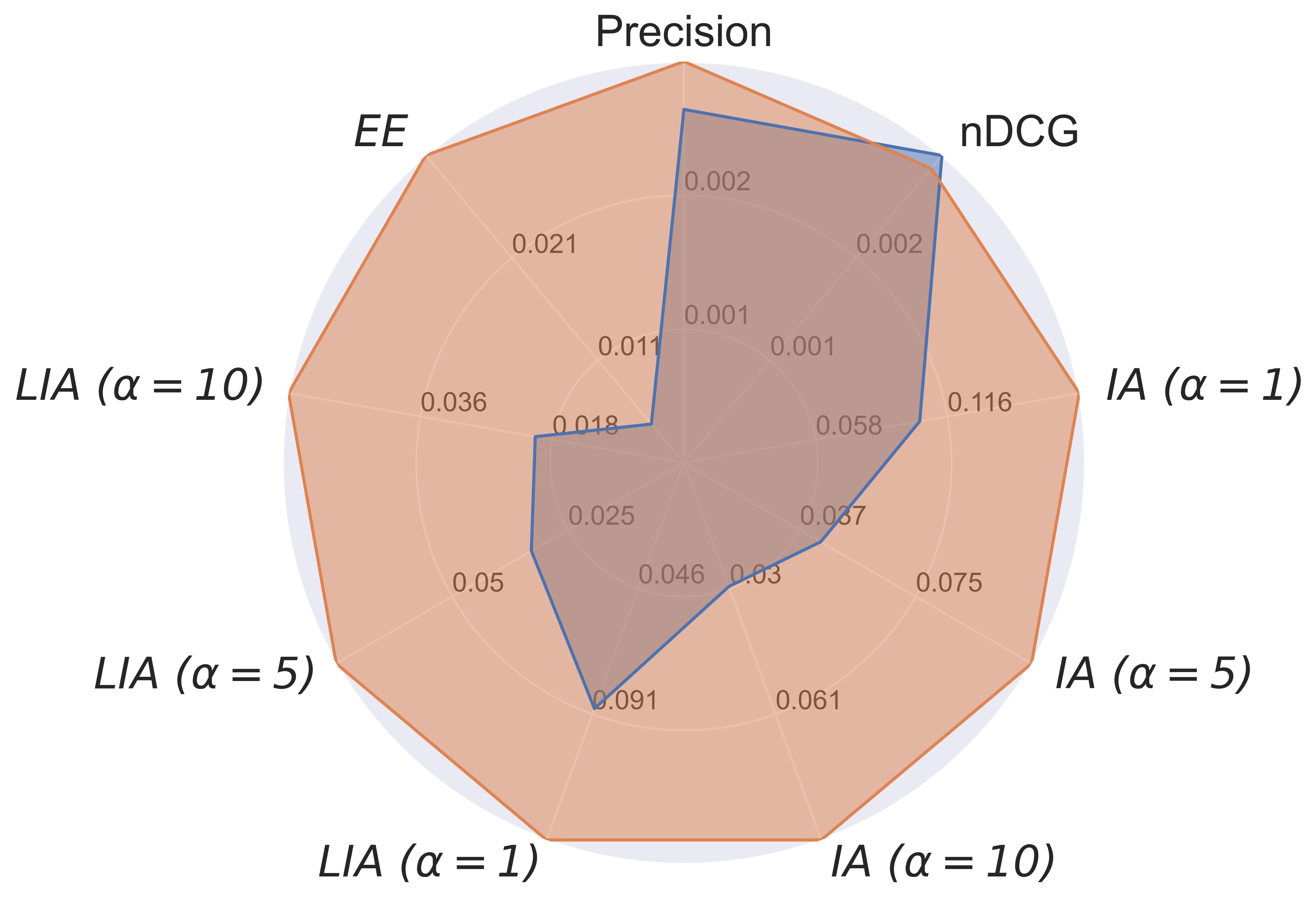}
        \caption{BiasedMF}
    \end{subfigure}
    \qquad
    \begin{subfigure}[b]{0.45\textwidth}
        \includegraphics[width=\textwidth]{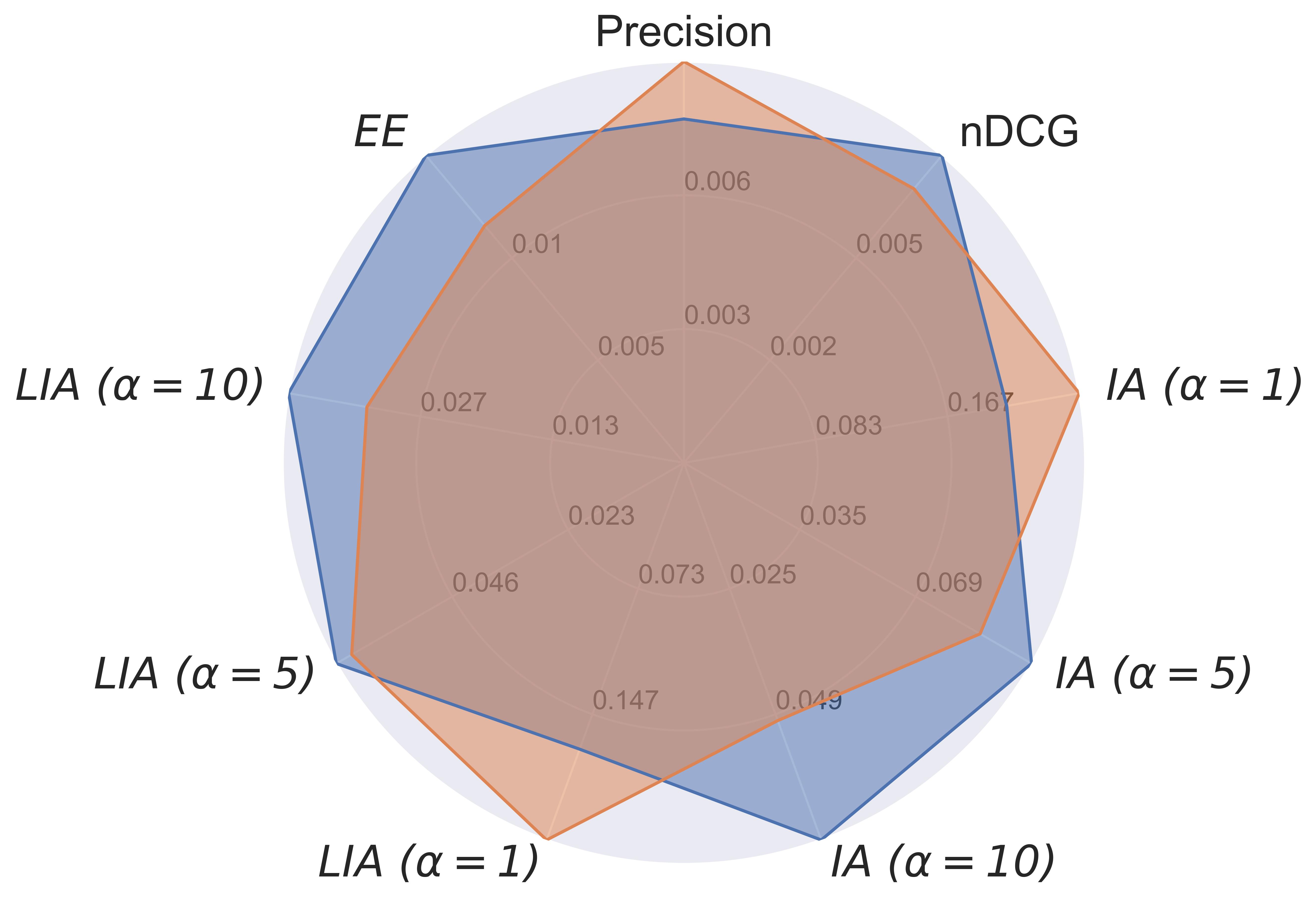}
        \caption{ListRankMF}
    \end{subfigure}
    \\
    \begin{subfigure}[b]{0.45\textwidth}
        \includegraphics[width=\textwidth]{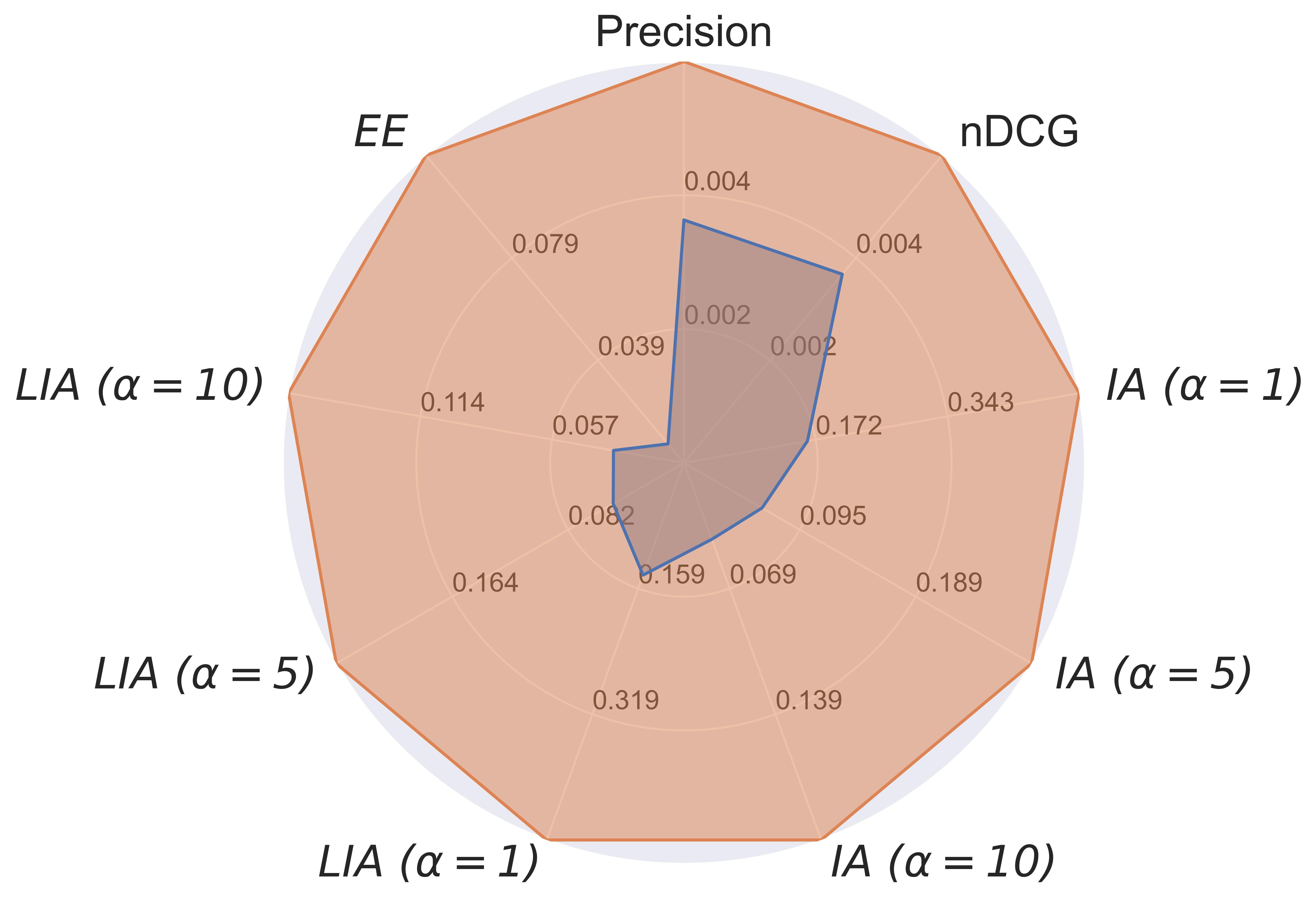}
        \caption{SVD++}
    \end{subfigure}
    \qquad
    \begin{subfigure}[b]{0.45\textwidth}
        \includegraphics[width=\textwidth]{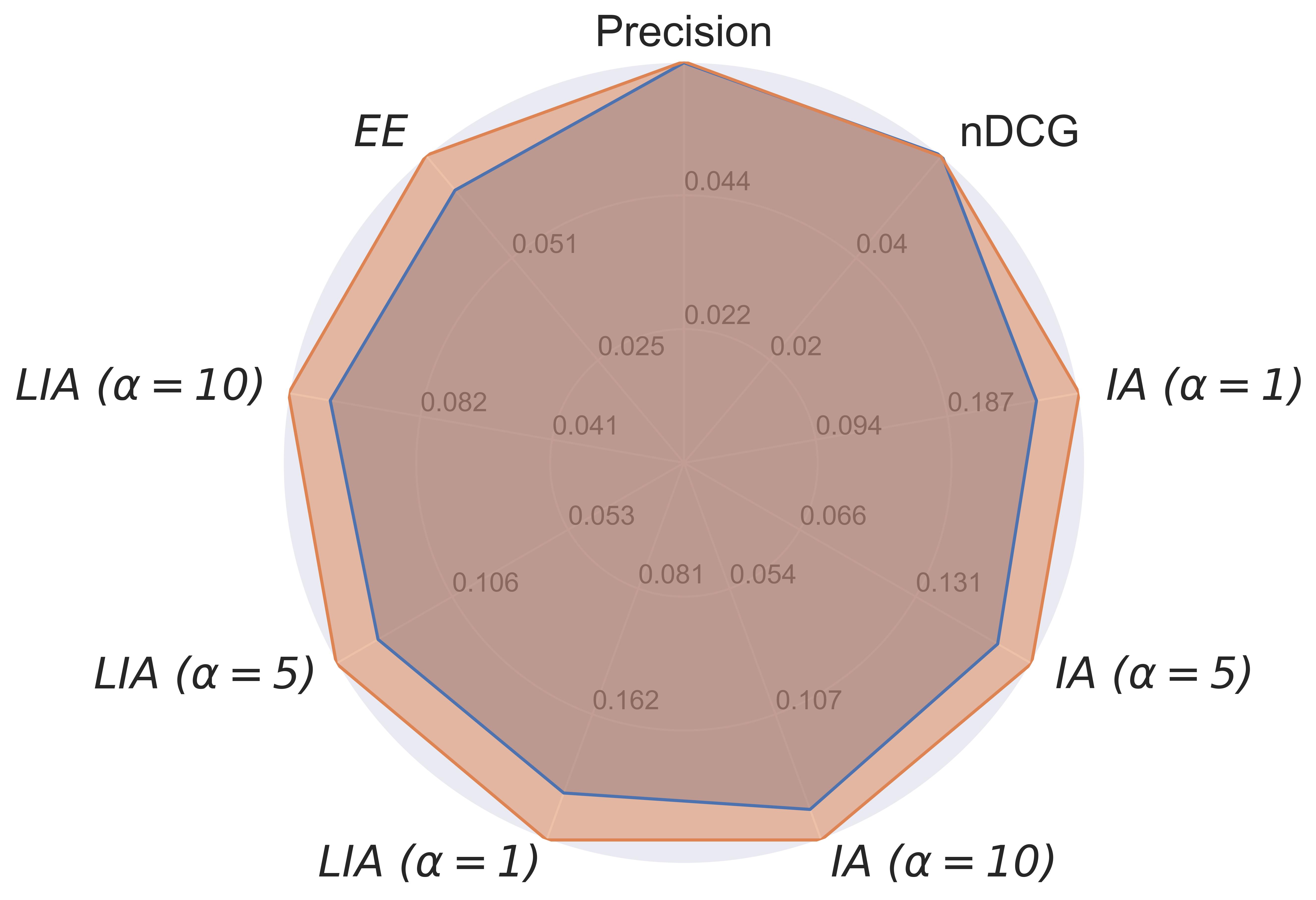}
        \caption{WRMF}
    \end{subfigure}
    \\
    \begin{subfigure}[b]{0.45\textwidth}
        \includegraphics[width=\textwidth]{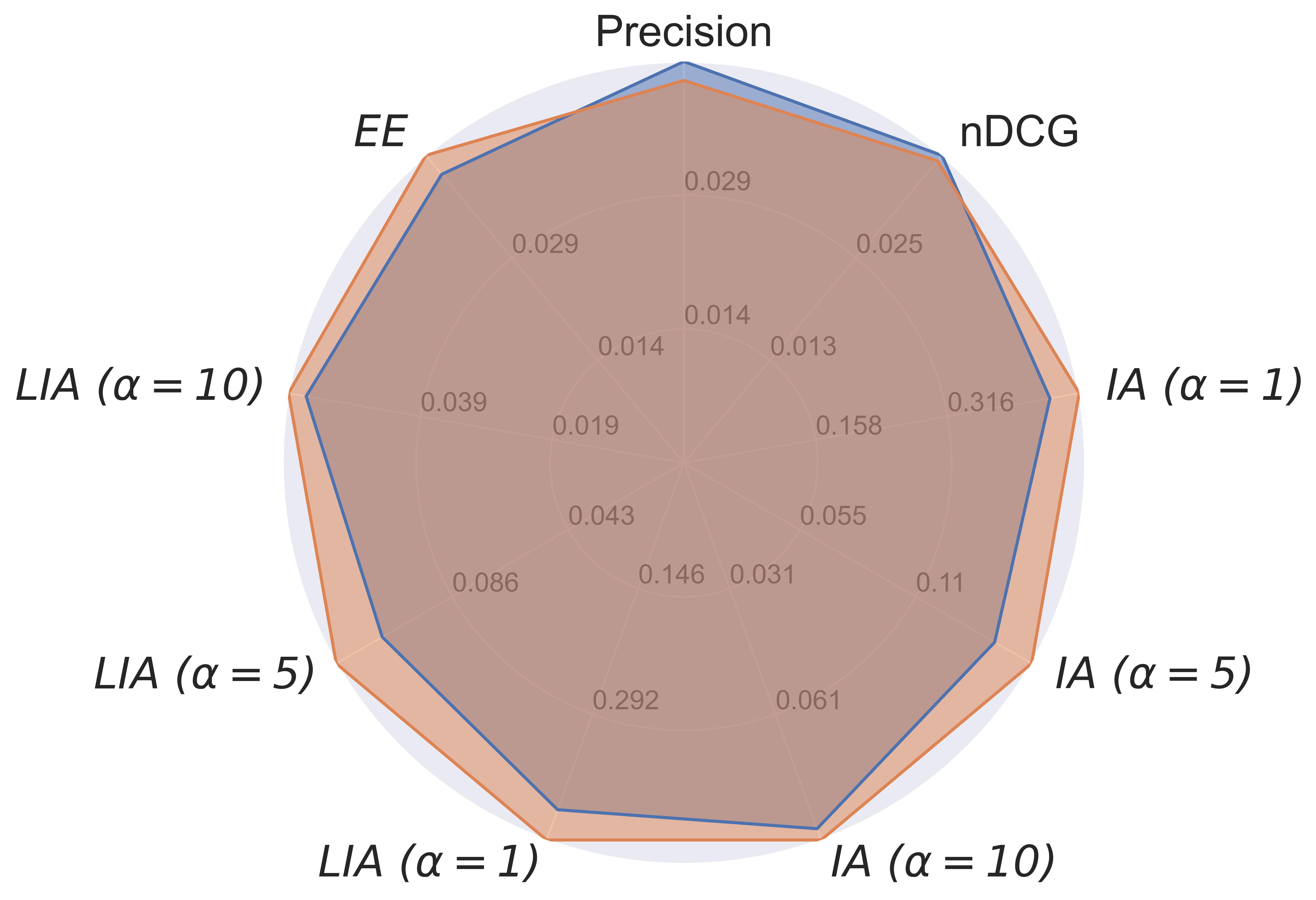}
        \caption{UserKNN}
    \end{subfigure}
    \qquad
    \begin{subfigure}[b]{0.45\textwidth}
        \includegraphics[width=\textwidth]{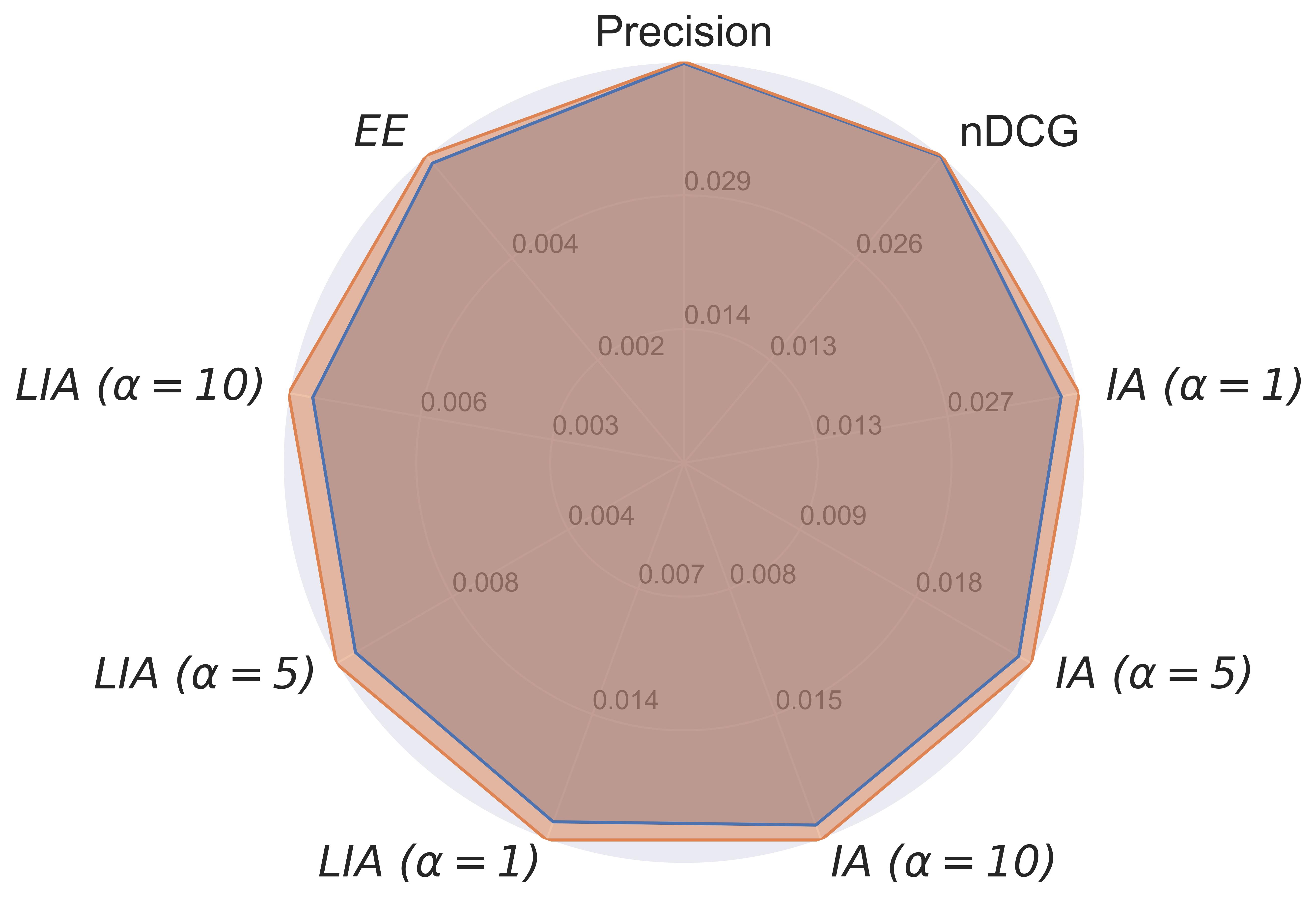}
        \caption{ItemKNN}
    \end{subfigure}
\caption{Performance of six recommendation algorithms when raw rating data and percentile data are separately used as input on the Yelp dataset. For all metrics, higher values signify a better performance.}\label{fig_yelp_pop_rating_sim}
\end{figure*}

Our hypothesis is that using this percentile matrix as input for recommendation algorithms results in a less biased model, that may consequently improve the fairness of recommendation results. To confirm this hypothesis, we separately perform experiments with raw rating data and transformed percentile data as input for the recommendation algorithms and evaluate their fairness. In the following section, we report our experimental results.

\subsection{Experimental results}

To show the effectiveness of the percentile transformation in improving fairness of exposure for items, we perform experiments using six recommendation models (described in Section~\ref{sec_algs}) on four datasets (Goodreads, MovieLens, Google Local Data, and Yelp). The recommendation models are separately trained with two types of input data: (i)~original rating data, and (ii)~percentile values. The results are evaluated in terms of both accuracy and fairness metrics (described in Section~\ref{sec_metrics}). Figures~\ref{fig_goodreads_pop_rating_sim}--\ref{fig_yelp_pop_rating_sim} present our experimental results.

First, from the plots in Figures~\ref{fig_goodreads_pop_rating_sim}--\ref{fig_yelp_pop_rating_sim} it is evident that our proposed percentile transformation significantly improves the fairness of exposure for items across all datasets, recommendation models, and evaluation metrics (except for \algname{ListRankMF} on the Yelp dataset). In particular, in most cases, not only fairness is significantly improved, but also accuracy metrics are boosted. For instance, on the Goodreads dataset, both accuracy and fairness are improved when percentile values are used as input for \algname{BiasedMF}, \algname{ListRankMF}, \algname{SVD++}, and \algname{ItemKNN}, rather than original rating values. This indicates the superiority of the proposed percentile transformation compared to existing unfairness mitigation methods that their fairness improvement is often at the expense of losing accuracy~\cite{greenwood2024user,ge2022toward,wang2022providing,liu2021balancing,xu2025understanding}. 

This fairness-accuracy trade-off can be also observed in our results, such as for \algname{WRMF} on the MovieLens dataset. However, the drop in accuracy is negligible in comparison to the gain obtained from the improvements in fairness. For example, for \algname{WRMF} on the Google Local Data, with a 1.22\% drop in nDCG, fairness is improved by 16.81\%, 19\%, and 18.16\% for $IA(\alpha=1)$, $LIA(\alpha=1)$, and $EE$, respectively. Or for \algname{UserKNN} on the MovieLens dataset, with a 7.61\% drop in nDCG, fairness is improved by 48.09\%, 85.04\%, and 76.18\% for $IA(\alpha=1)$, $LIA(\alpha=1)$, and $EE$, respectively. This confirms the effectiveness of the proposed solution in achieving exposure fairness for items.

For \algname{UserKNN} on the Goodreads dataset, the results show that our percentile method is outperformed by the original model in terms of $IA (\alpha=10)$ and $LIA (\alpha=10)$. This is due to the fact that our percentile method recommends more unique items in recommendation lists (higher $IA (\alpha=1)$ and $LIA (\alpha=1)$), but there are not sufficiently many recommendation slots to show each at least 10 times (only 10 slots per user are available). This results in lower $IA (\alpha=10)$ and $LIA (\alpha=10)$ values, while higher $IA (\alpha=1)$, $LIA (\alpha=1)$, and $\mathit{EE}$ values signify that more unique items are equally represented in the recommendation results, which follows the fairness criteria. 
\section{Comparing with Baselines and Improving Existing Unfairness Mitigation Pipelines}

In this section, we further investigate how our proposed pre-processing method can enhance the performance of existing fairness mitigation strategies. As demonstrated in Section~\ref{sec_percentile}, our percentile-based pre-processing method significantly improves the fairness of recommendation models while preserving their accuracy. Next, we extend that analysis by examining the effectiveness of our pre-processing method when combined with existing fairness mitigation approaches, particularly within the broader recommendation pipeline.

As discussed in Section~\ref{sec_relatedwork}, existing fairness mitigation techniques are commonly categorized into three groups: pre-processing, in-processing, and post-processing. Among these, post-processing methods, also known as rerankers, have received considerable attention due to two key advantages:
(i) they are model-agnostic and thus applicable across various recommendation models, and
(ii) they have demonstrated strong performance in fairness enhancement, owing to their flexibility in optimizing fairness criteria.

In a typical post-processing approach, an initial (often long) recommendation list is generated for each user by a base recommendation model. The reranker then refines this list to produce a shorter and fairer top-$K$ recommendation output. However, it has been shown that the size of the initial recommendation list directly affects both fairness outcomes and computational efficiency. Longer initial lists generally lead to higher fairness but require greater computational resources and time. Shorter initial lists are more efficient but often result in reduced fairness~\cite{zehlike2017fa,mansoury2020fairmatch,antikacioglu2017post}. 

Our central hypothesis is that if a recommendation model can produce sufficiently fair and accurate initial recommendation lists—even with a smaller size—then the reranker has a better foundation to work with, potentially achieving strong fairness outcomes at reduced computational cost. To demonstrate this, we evaluate the effectiveness of post-processing methods when applied to initial recommendation lists generated using models trained on percentile values (i.e., after applying our pre-processing method). 

For this analysis, we focus exclusively on our percentile-based transformation, not alternative pre-processing methods, for two reasons. First, to the best of our knowledge, no existing pre-processing fairness method operates by modifying rating values to mitigate positivity bias. While several pre-processing fairness approaches exist~\cite{oldfield2025revisiting,waris2024novel,rus2024study,chen2021counterfactual,lahoti2019ifair,sonoda2021pre}, none are designed to adjust the rating distribution itself, which is essential for addressing multifactorial bias. Second, introducing a new pre-processing algorithm aimed at optimally improving fairness is not the primary objective of this work. Rather, our goal is to demonstrate the role that multifactorial bias~\cite{huang2024going} plays in shaping item-side exposure fairness and to show that even a simple, data-driven transformation can meaningfully reduce this bias in the input data. The percentile transformation serves as a clear and effective example of how mitigating multifactorial bias at the data level can lead to improved item exposure fairness in downstream recommendation outputs.

In our experiments, we use \algname{BiasedMF} as the base recommendation model. We evaluate the following four reranking algorithms: Discrepancy Minimization (DM)~\cite{antikacioglu2017post}, FA*IR~\cite{zehlike2017fa}, xQuAD~\cite{abdollahpouri2019managing}, and FairMatch~\cite{mansoury2020fairmatch}. These rerankers are described in Section~\ref{sec_relatedwork}. Additionally, we include two naïve baselines for comparison: 
(i)~Random, which randomly selects $K$ items from the initial list, and (
(ii)~Reverse, which selects the bottom-$K$ items based on their original relevance scores (i.e., sorted in ascending order).

We test three initial recommendation list sizes: ${20, 50, 100}$, from which we generate a final top-10 recommendation list using the rerankers. We evaluate four experimental settings:

\begin{itemize}
	\item Original ratings only: $rating \rightarrow \text{BiasedMF}$
	\item Percentile values only: $percentile \rightarrow \text{BiasedMF}$
	\item Reranking applied on original ratings: $rating \rightarrow \text{BiasedMF} \rightarrow \text{reranker}$
	\item Reranking applied on percentile values: $percentile \rightarrow \text{BiasedMF} \rightarrow \text{reranker}$
\end{itemize}

\noindent%
Figures~\ref{fig_ndcg_ia} and~\ref{fig_ndcg_equality} present results across four datasets using six rerankers, plotted in terms of nDCG vs. $IA (\alpha=1)$ and nDCG vs. $EE$, respectively.\footnote{Similar trends were observed across other metrics as well.} Several interesting patterns can be observed in these plots.

\input{figures/fig_ndcg_ia}

\input{figures/fig_ndcg_equality}

First, we observe a clear trade-off between the size of the initial recommendation list and the resulting performance. Larger initial lists (e.g., size 100) consistently lead to greater fairness, but at the cost of lower accuracy. Smaller initial lists (e.g., size 20) maintain higher accuracy but deliver less fairness. This trade-off is well-documented in prior work~\cite{mehrotra2018towards,ge2022toward}. Our results align with these findings.

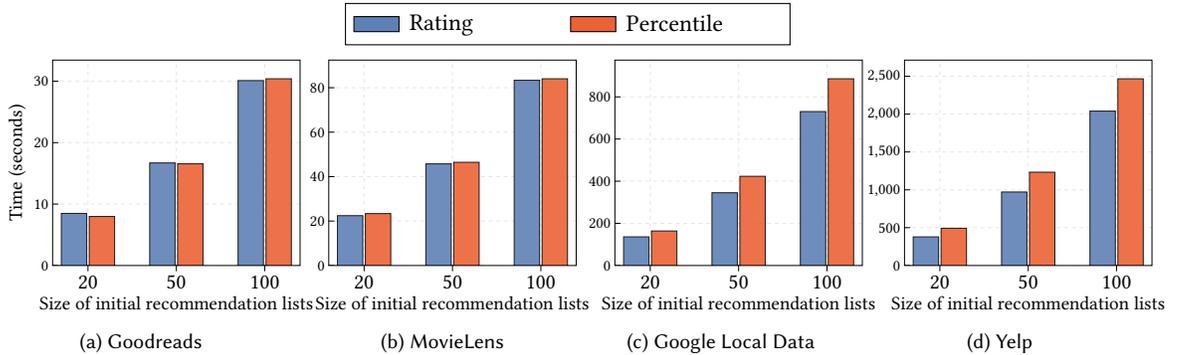
\begin{figure}

        \centering
        \begin{subfigure}[b]{\textwidth}
        \centering
        \begin{tikzpicture}    
            \begin{axis}[
                hide axis,
                xmin=0, xmax=1, ymin=0, ymax=1,
                legend to name=mylegend,
                legend columns=2,
                legend style={
                    font=\normalsize,
                    text=black,  
                    column sep=0.2em,
                    draw=black,        
                    cells={anchor=west},
                    text width=2cm,    
                    at={(0.5,1.05)}, 
                    anchor=south,
                },
                legend image code/.code={
                    \draw[fill=#1, draw=black, line width=0.5pt] 
                        (0cm,-0.1cm) rectangle (0.6cm,0.2cm);
                },
            ]
            \addlegendimage{area legend, fill=tolblue} 
            \addlegendentry{Rating}
            \addlegendimage{area legend, fill=tolorange} 
            \addlegendentry{Percentile}
            \end{axis}
        \end{tikzpicture}
        \end{subfigure}

        \vspace{-15em}
        \ref{mylegend}
        \vspace{0.5em}

        \begin{subfigure}[b]{0.24\textwidth}
        \centering
        \begin{tikzpicture}[scale=0.48]
            \begin{axis}[
                ybar,
                bar width=20pt,
                ymin=0,
                enlarge x limits=0.2,
                axis lines=box,
                xtick pos=left,
                ytick pos=left,
                tick style={line cap=round, black},
                xtick=data,
                ylabel={Time (seconds)},
                xlabel={Size of initial recommendation lists},
                symbolic x coords={20, 50, 100},
                tick label style={font=\large},
                label style={font=\fontsize{15}{20}\selectfont},
                grid=major,
                grid style={dashed,gray!20},
                nodes near coords={},
                xticklabel style={/pgf/number format/fixed, font=\fontsize{17}{20}\selectfont},
                yticklabel style={/pgf/number format/fixed, font=\fontsize{12}{20}\selectfont},
            ]
            \addplot+[
                ybar,
                fill=tolblue,
                draw=black,
                opacity=0.9
            ] coordinates {(20,8.484023) (50,16.696577) (100,30.095961)};
            
            \addplot+[
                ybar,
                fill=tolorange,
                draw=black,
                opacity=0.9
            ] coordinates {(20,7.979232) (50,16.55477) (100,30.391467)};
            
            \end{axis}
        \end{tikzpicture}
        \caption{Goodreads}
        \end{subfigure}
\hspace{3mm}
        \begin{subfigure}[b]{0.24\textwidth}
        \centering
        \begin{tikzpicture}[scale=0.48]
            \begin{axis}[
                ybar,
                bar width=20pt,
                ymin=0,
                enlarge x limits=0.2,
                axis lines=box,
                xtick pos=left,
                ytick pos=left,
                tick style={line cap=round, black},
                xtick=data,
                xlabel={Size of initial recommendation lists},
                symbolic x coords={20, 50, 100},
                tick label style={font=\large},
                label style={font=\fontsize{15}{20}\selectfont},
                grid=major,
                grid style={dashed,gray!20},
                nodes near coords={},
                xticklabel style={/pgf/number format/fixed, font=\fontsize{17}{20}\selectfont},
                yticklabel style={/pgf/number format/fixed, font=\fontsize{12}{20}\selectfont},
            ]
            \addplot+[
                ybar,
                fill=tolblue,
                draw=black,
                opacity=0.9
            ] coordinates {(20,22.422019) (50,45.745912) (100,83.344932)};
            
            \addplot+[
                ybar,
                fill=tolorange,
                draw=black,
                opacity=0.9
            ] coordinates {(20,23.364973) (50,46.443438) (100,84.024392)};
            \end{axis}
        \end{tikzpicture}
        \caption{MovieLens}
        \end{subfigure}
\hspace{0mm}
        \begin{subfigure}[b]{0.24\textwidth}
        \centering
        \begin{tikzpicture}[scale=0.48]
            \begin{axis}[
                ybar,
                bar width=20pt,
                ymin=0,
                enlarge x limits=0.2,
                axis lines=box,
                xtick pos=left,
                ytick pos=left,
                tick style={line cap=round, black},
                xtick=data,
                xlabel={Size of initial recommendation lists},
                symbolic x coords={20, 50, 100},
                tick label style={font=\large},
                label style={font=\fontsize{15}{20}\selectfont},
                grid=major,
                grid style={dashed,gray!20},
                nodes near coords={},
                xticklabel style={/pgf/number format/fixed, font=\fontsize{17}{20}\selectfont},
                yticklabel style={/pgf/number format/fixed, font=\fontsize{12}{20}\selectfont},
            ]
            \addplot+[
                ybar,
                fill=tolblue,
                draw=black,
                opacity=0.9
            ] coordinates {(20,136.012769) (50,345.116806) (100,730.207654)};
            
            \addplot+[
                ybar,
                fill=tolorange,
                draw=black,
                opacity=0.9
            ] coordinates {(20,163.795041) (50,423.234219) (100,885.815593)};
            \end{axis}
        \end{tikzpicture}
        \caption{Google Local Data}
        \end{subfigure}
\hspace{0mm}
        \begin{subfigure}[b]{0.24\textwidth}
        \centering
        \begin{tikzpicture}[scale=0.48]
            \begin{axis}[
                ybar,
                bar width=20pt,
                ymin=0,
                enlarge x limits=0.2,
                axis lines=box,
                xtick pos=left,
                ytick pos=left,
                tick style={line cap=round, black},
                xtick=data,
                xlabel={Size of initial recommendation lists},
                symbolic x coords={20, 50, 100},
                tick label style={font=\large},
                label style={font=\fontsize{15}{20}\selectfont},
                grid=major,
                grid style={dashed,gray!20},
                nodes near coords={},
                xticklabel style={/pgf/number format/fixed, font=\fontsize{17}{20}\selectfont},
                yticklabel style={/pgf/number format/fixed, font=\fontsize{12}{20}\selectfont},
            ]
            \addplot+[
                ybar,
                fill=tolblue,
                draw=black,
                opacity=0.9
            ] coordinates {(20,378.09987) (50,971.19437) (100,2039.69206)};
            
            \addplot+[
                ybar,
                fill=tolorange,
                draw=black,
                opacity=0.9
            ] coordinates {(20,492.082716) (50,1232.870531) (100,2466.79068)};
            \end{axis}
        \end{tikzpicture}
        \caption{Yelp}
        \end{subfigure}
    \caption{Running time of DM reranker with both original and percentile input types on four datasets.}
    \label{fig_time}
\end{figure}

Second, reranking methods require increased processing time as the initial list grows. Figure~\ref{fig_time} shows the runtime of DM across different initial list sizes and datasets. For example, on the MovieLens dataset, DM takes approximately 3.7 times longer to process an initial list of size 100 compared to size 20, regardless of whether ratings or percentile values are used as input. Ideally, we aim to achieve high fairness and low accuracy loss without the need for large initial lists, which leads to the next observation.

Third, our percentile transformation method clearly meets the ideal of achieving high fairness with minimal accuracy loss and reduced computational overhead. Across all datasets, rerankers, and evaluation metrics, using percentile-transformed input consistently results in: (i) higher fairness scores, (ii) comparable or improved accuracy, and (iii) notably strong performance even with smaller initial recommendation lists.

For example, as shown in Figure~\ref{fig_ndcg_ia} for the MovieLens dataset, the FairMatch reranker applied to percentile-transformed input with an initial list size of 20 achieved an $IA (\alpha=1)$ of 0.335 and an nDCG of 0.11, substantially outperforming the same reranker using an initial list of size 100 trained on the original ratings, which yielded $IA (\alpha=1)$ of 0.253 and nDCG of 0.068.

Similarly, in Figure~\ref{fig_ndcg_equality}, the same reranker with percentile-transformed input achieved an $\mathit{EE}$ score of 0.05 on an initial list of size 20, compared to just 0.006 when using original rating input with a larger list size of 100.

To further understand the performance boost provided by our method, we compare the utility of DM reranker when applied to: (i) shorter list (size 20) generated using percentile values, and (ii) longer list (size 100) generated using original rating values. We compute the relative gain as:
\begin{equation}
\mathit{gain}^{\mathcal{M}} = \frac{\mathit{Utility}^{\mathcal{M},20}_{\mathit{Percentile}} - \mathit{Utility}^{\mathcal{M},100}_{\mathit{Rating}}}{\mathit{Utility}^{\mathcal{M},100}_{\mathit{Rating}}} \times 100,
\end{equation}
where $\mathit{Utility}^{\mathcal{M},20}_{\mathit{Percentile}}$ denotes the performance score (e.g., nDCG, IA, $\mathit{EE}$) of the final recommendation list produced by DM using percentile input and a 20-item initial list. $\mathit{Utility}^{\mathcal{M},100}_{\mathit{Rating}}$ represents the same score using original ratings and a 100-item initial list. This analysis allows us to assess whether our method can match or exceed the performance of traditional pipelines while using far fewer resources.

Table~\ref{tab_gain} presents the gain values for each dataset and metric. The results strongly favor our method across nearly all cases. For instance, we observe a 207.84\% nDCG gain on the Google Local Data and a substantial $EE$ gain on the Yelp dataset (109.11\%). An exception is found only in the $\mathit{EE}$ metric on the Google Local Data, where the performance is comparable. These results demonstrate that using percentile values as input enables recommendation models to generate fairer initial lists, which in turn allows rerankers to achieve better final fairness and accuracy outcomes, even with shorter input lists, leading to a more effective and efficient fairness-aware recommendation pipeline.

\begin{table}[t]
\centering
\captionof{table}{Performance gain of DM reranker with initial recommendation lists of size 20 generated by percentile values compared to initial recommendation list of size 100 generated by rating values on four datasets. } \label{tab_gain}
\begin{tabular}{lrrrr}
\toprule
 Dataset & nDCG & IA ($\alpha=1$) & EE & Running time \\
 \midrule

Goodreads & 57.69\% & 30.64\% & 20.25\% & 73.49\% \\ 
ML & 79.52\% & 35.79\% & 83.78\% & 71.97\% \\
Google Local Data & 207.84\% & 7.30\% & $-$11.91\% & 77.57\% \\
Yelp & 37.65\% & 61.98\% & 109.11\% & 75.87\% \\

\bottomrule
\end{tabular}
\end{table}
\section{Conclusion}

In this paper, we have investigated the combined effect of popularity bias and positivity bias, to which we refer as multifactorial bias, on the fairness of recommender systems. Our analysis revealed that positivity bias disproportionately affects popular items, intensifying their over-exposure and leading to unfair treatment of less popular items in recommendation results. We demonstrated through simulation and empirical studies that multifactorial bias plays a significant role in amplifying exposure bias, a critical form of item-side unfairness.

To mitigate this issue, we adapted a percentile-based rating transformation as a pre-processing method aimed at reducing multifactorial bias in the input data. Our extensive experiments across multiple datasets and recommendation algorithms showed that this approach improves exposure fairness without compromising accuracy. Moreover, we showed that incorporating this method into existing post-processing fairness pipelines can enhance both their effectiveness and efficiency, enabling fair outcomes with shorter initial recommendation lists and lower computational cost.

Overall, these findings highlight the need to explicitly address multifactorial bias in recommender systems and demonstrate that simple, data-driven pre-processing techniques can play a powerful role in building fairer and more efficient recommendation pipelines.

Despite these contributions, the percentile-based transformation method has two key limitations. First, its reliability depends on the availability of a sufficient number of rating values; for example, applying a percentile transformation to a rating vector such as $\langle 3,4 \rangle$ would be unstable. Second, the transformation assumes diversity in rating values; a uniform vector such as $\langle 3,3,\ldots,3 \rangle$ would yield uninformative percentile scores. In this work, we focused on dense datasets to avoid these issues, but a natural direction for future research is to examine how percentile-based transformations behave under sparse data conditions. Given that data sparsity constitutes a fundamental challenge in recommender systems, future research should examine the robustness and generalizability of our approach in sparse data environments and cold-start scenarios.

Additionally, future work will also explore whether combining existing mitigation strategies for popularity and positivity biases can effectively address multifactorial bias when applied together. We further aim to extend our study to user-side fairness, investigating how alleviating multifactorial bias in the data may improve fairness in recommendations delivered to users.

\begin{acks}
This research was (partially) supported by the Dutch Research Council (NWO), under project numbers 024.004.022, NWA.1389.20.\-183, and KICH3.LTP.20.006, and the European Union under grant agreements No. 101070212 (FINDHR) and No. 1012
01510 (UNITE).

Views and opinions expressed are those of the author(s) only and do not necessarily reflect those of their respective employers, funders and/or granting authorities.
\end{acks}

\section*{Resources}
To facilitate the reproducibility of experiments in this paper, we make the code and datasets available at:
\url{https://github.com/masoudmansoury/Unfairness_MultifactorialBias}.

\bibliographystyle{ACM-Reference-Format}
\bibliography{references}

\appendix

\end{document}